\documentclass[twocolumn]{aastex62}
\usepackage{rotating}
\usepackage{color, colortbl}
\usepackage{tikz}
 
\usepackage{pifont}
\newcommand{\cmark}{\color{green} \ding{51}}%
\newcommand{\xmark}{\color{red}\ding{56}}%
\newcommand{\qmark}{\color{black}{\bf ?}}%
\usepackage[para]{threeparttable}

\definecolor{palecopper}{rgb}{0.85, 0.54, 0.4}
\definecolor{palegold}{rgb}{0.9, 0.75, 0.54}
\definecolor{papayawhip}{rgb}{1.0, 0.94, 0.84}
\definecolor{papayawhip}{rgb}{1.0, 0.94, 0.84}
\definecolor{cadmiumgreen}{rgb}{0.0, 0.42, 0.24}

\graphicspath{{./}{figures/}}

\submitjournal{ApJ}

\shorttitle{ZTF Flash Spectroscopy}
\shortauthors{Bruch et al.}

\begin{document}

\title{A large fraction of hydrogen-rich supernova progenitors experience elevated mass loss shortly prior to explosion}

\correspondingauthor{Rachel J. Bruch}
\email{rachel.bruch@weizmann.ac.il}

\author[0000-0002-0786-7307]{Rachel J. Bruch}
\affil{Department of Particle Physics and Astrophysics 
Weizmann Institute of Science 
234 Herzl St.
76100 Rehovot, Israel}

\author{Avishay Gal-Yam}
\affil{Department of Particle Physics and Astrophysics 
Weizmann Institute of Science 
234 Herzl St.
76100 Rehovot, Israel}

\author[0000-0001-6797-1889]{Steve Schulze}
\affil{Department of Particle Physics and Astrophysics 
Weizmann Institute of Science 
234 Herzl St.
76100 Rehovot, Israel}

\author{Ofer Yaron}
\affil{Department of Particle Physics and Astrophysics 
Weizmann Institute of Science 
234 Herzl St.
76100 Rehovot, Israel}

\author{Yi Yang}
\affil{Department of Particle Physics and Astrophysics 
Weizmann Institute of Science 
234 Herzl St.
76100 Rehovot, Israel}

\author[0000-0001-6753-1488]{Maayane Soumagnac}
\affil{Department of Particle Physics and Astrophysics 
Weizmann Institute of Science 
234 Herzl St.
76100 Rehovot, Israel}
\affil{Computational Cosmology Center, Lawrence Berkeley National Laboratory, 1 Cyclotron Road, Berkeley, CA 94720, USA}

\author{Mickael Rigault}
\affil{Universit\'e de Lyon, Universit\'e Claude Bernard Lyon 1, CNRS/IN2P3, IP2I Lyon, F-69622, Villeurbanne, France}

\author[0000-0002-4667-6730]{Nora L. Strotjohann}
\affil{Department of Particle Physics and Astrophysics 
Weizmann Institute of Science 
234 Herzl St.
76100 Rehovot, Israel}

\author{Eran Ofek}
\affil{Department of Particle Physics and Astrophysics 
Weizmann Institute of Science 
234 Herzl St.
76100 Rehovot, Israel}

\author{Jesper Sollerman}
\affil{The Oskar Klein Centre, Department of Astronomy, Stockholm University, AlbaNova, SE-106 91 Stockholm, Sweden}

\author[0000-0002-8532-9395]{Frank J. Masci}
\affiliation{IPAC, California Institute of Technology, 1200 E. California Blvd, Pasadena, CA 91125, USA}

\author[0000-0002-3821-6144]{Cristina Barbarino}
\affil{The Oskar Klein Centre, Department of Astronomy, Stockholm University, AlbaNova, SE-106 91 Stockholm, Sweden}

\author[0000-0002-9017-3567]{Anna Y. Q.~Ho}
\affiliation{Cahill Center for Astrophysics, 
California Institute of Technology, MC 249-17, 
1200 E California Boulevard, Pasadena, CA, 91125, USA}

\author[0000-0002-4223-103X]{Christoffer Fremling}
\affil{Cahill Center for Astrophysics, California Institute of Technology, MC 249-17, 1200 E California Boulevard, Pasadena, CA, 91125, USA}

\author[0000-0001-8472-1996]{Daniel Perley}
\affil{Astrophysics Research Institute, Liverpool John Moores University, Liverpool Science Park, 146 Brownlow Hill, Liverpool L3 5RF, UK}

\author{Jakob Nordin}
\affil{Institute of Physics, Humboldt-Universit¨at zu Berlin, Newtonstr. 15, 12489 Berlin, Germany}

\author{S. Bradley Cenko}
\affil{Astrophysics Science Division, NASA Goddard Space Flight
Center, MC 661, Greenbelt, MD 20771, USA}
\affil{Joint Space-Science Institute, University of Maryland, College Park, MD 20742, USA}

\author{S. Adams}
\affil{Cahill Center for Astrophysics, California Institute of Technology, MC 249-17, 1200 E California Boulevard, Pasadena, CA, 91125, USA}

\author{Igor Adreoni}
\affil{Cahill Center for Astrophysics, California Institute of Technology, MC 249-17, 1200 E California Boulevard, Pasadena, CA, 91125, USA}

\author[0000-0001-8018-5348]{Eric C. Bellm}
\affil{DIRAC Institute, Department of Astronomy, University of Washington, 3910 15th Avenue NE, Seattle, WA 98195, USA}

\author[0000-0003-0901-1606]{Nadia Blagorodnova}
\affil{Department of Astrophysics/IMAPP, Radboud University, Nijmegen, The Netherlands}

\author[0000-0002-8255-5127]{Mattia Bulla}
\affil{The Oskar Klein Centre, Department of Astronomy, Stockholm University, AlbaNova, SE-106 91 Stockholm, Sweden}

\author{Kevin Burdge}
\affil{Cahill Center for Astrophysics, California Institute of Technology, MC 249-17, 1200 E California Boulevard, Pasadena, CA, 91125, USA}

\author[0000-0002-8989-0542]{Kishalay De}
\affil{Cahill Center for Astrophysics, California Institute of Technology, MC 249-17, 1200 E California Boulevard, Pasadena, CA, 91125, USA}

\author{Suhail Dhawan}
\affil{The Oskar Klein Centre, Department of Astronomy, Stockholm University, AlbaNova, SE-106 91 Stockholm, Sweden}

\author{Andrew J. Drake}
\affil{Division of Physics, Mathematics and Astronomy, California Institute of Technology, Pasadena, CA 91125, USA}

\author[0000-0001-5060-8733]{Dmitry A. Duev}
\affil{Division of Physics, Mathematics and Astronomy, California Institute of Technology, Pasadena, CA 91125, USA}

\author{Alison Dugas}
\affil{Department of Physics and Astronomy, Watanabe 416, 2505 Correa Road, Honolulu, HI 96822}

\author[0000-0002-3168-0139]{Matthew Graham}
\affil{Cahill Center for Astrophysics, California Institute of Technology, MC 249-17, 1200 E California Boulevard, Pasadena, CA, 91125, USA}

\author{Melissa L. Graham}
\affil{University of Washington, Department of Astronomy Box 351580 Seattle WA 98195-1580, USA}

\author{Jacob Jencson}
\affil{Cahill Center for Astrophysics, California Institute of Technology, MC 249-17, 1200 E California Boulevard, Pasadena, CA, 91125, USA}

\author{Emir Karamehmetoglu}
\affil{The Oskar Klein Centre, Department of Astronomy, Stockholm University, AlbaNova, SE-106 91 Stockholm, Sweden}
\affil{Department of Physics and Astronomy, Aarhus University, Ny Munkegade 120, DK-8000 Aarhus C, Denmark}

\author[0000-0002-5619-4938]{Mansi Kasliwal}
\affil{Cahill Center for Astrophysics, California Institute of Technology, MC 249-17, 1200 E California Boulevard, Pasadena, CA, 91125, USA}

\author[0000-0002-1031-0796]{Young-Lo Kim}
\affil{Universit\'e de Lyon, Universit\'e Claude Bernard Lyon 1, CNRS/IN2P3, IP2I Lyon, F-69622, Villeurbanne, France}

\author[0000-0001-5390-8563]{Shrinivas Kulkarni}
\affil{Cahill Center for Astrophysics, California Institute of Technology, MC 249-17, 1200 E California Boulevard, Pasadena, CA, 91125, USA}

\author[0000-0002-6540-1484]{Thomas Kupfer}
\affil{Kavli Institute for Theoretical Physics, University of California, Santa Barbara, CA 93106, USA}

\author{Ashish Mahabal}
\affil{Division of Physics, Mathematics, and Astronomy, California Institute of Technology, Pasadena, CA 91125, USA}

\author[0000-0001-9515-478X]{A.~A.~Miller}
\affiliation{Center for Interdisciplinary Exploration and Research in Astrophysics and Department of Physics and Astronomy, Northwestern University, 1800 Sherman Ave, Evanston, IL 60201, USA}
\affiliation{The Adler Planetarium, Chicago, IL 60605, USA}

\author{Thomas A. Prince}
\affil{Division of Physics, Mathematics and Astronomy, California Institute of Technology, Pasadena, CA 91125, USA}

\author{Reed Riddle}
\affil{Caltech Optical Observatories, California Institute of Technology, Pasadena, CA 91125, USA}

\author{Y. Sharma}
\affil{Cahill Center for Astrophysics, California Institute of Technology, MC 249-17, 1200 E California Boulevard, Pasadena, CA, 91125, USA}

\author{Roger Smith}
\affil{Caltech Optical Observatories, California Institute of Technology, Pasadena, CA 91125, USA}

\author{Francesco Taddia}
\affil{The Oskar Klein Centre, Department of Astronomy, Stockholm University, AlbaNova, SE-106 91 Stockholm, Sweden}
\affil{Department of Physics and Astronomy, Aarhus University, Ny Munkegade 120, DK-8000 Aarhus C, Denmark}

\author[0000-0002-5748-4558]{Kirsty Taggart}
\affil{Astrophysics Research Institute, Liverpool John Moores University, Liverpool Science Park, 146 Brownlow Hill, Liverpool L3 5RF, UK}

\author{Richard Walters}
\affil{Caltech Optical Observatories, California Institute of Technology, MC 249-17, 1200 E California Boulevard, Pasadena, CA, 91125}

\author{Lin Yan}
\affil{Cahill Center for Astrophysics, California Institute of Technology, MC 249-17, 1200 E California Boulevard, Pasadena, CA, 91125, USA}

\begin{abstract}
Spectroscopic detection of narrow emission lines traces the presence of circumstellar mass distributions around massive stars exploding as core-collapse supernovae. Transient emission lines disappearing shortly after the supernova explosion suggest that the spatial extent of such material is compact, and hence imply an increased mass loss shortly prior to explosion. Here, we present a systematic survey for such transient emission lines (Flash Spectroscopy) among Type II supernovae detected in the first year of the Zwicky Transient Facility (ZTF) survey. We find that at least six out of ten events for which a spectrum was obtained within two days of estimated explosion time show evidence for such transient flash lines. Our measured flash event fraction ($>30\%$ at $95\%$ confidence level) indicates that elevated mass loss is a common process occurring in massive stars that are about to explode as supernovae.  
\end{abstract}

\keywords{supernova:general - methods: observational - stars: mass-loss - stars: massive}

\section{Introduction}

Massive stars ($M>$ 8 M$_{\odot}$) explode as core-collapse supernovae (CC SNe; \citealt{smartt2015}; \citealt{galyam2017}). Such massive stars often experience mass loss from their outer layers, due to stellar winds, binary interaction, or eruptive mass-loss events (see, e.g., \citealt{smith2014} and references within). The mass lost by these stars forms distributions of circumstellar medium (CSM). The properties of the CSM depend on the mass-loss rate, the velocity of the flow, and the duration of the process.

When a massive star surrounded by CSM explodes as a CC SN, signatures of the CSM may manifest as spectroscopic features with a narrow width reflecting the mass-loss velocity, that is typically low compared to the expansion velocity of the supernova ejecta. In Type IIn SNe (e.g., \citealt{schlegel1990}, \citealt{filippenko1997}, \citealt{galyam2017}, \citealt{kiewe2012}, \citealt{taddia2013}, \citealt{nyholm2019}) narrow hydrogen lines persist for weeks to years after explosion, indicating an extensive CSM distribution. Type Ibn events (e.g., \citealt{pastorello2016}, \citealt{galyam2017}, \citealt{hossenzeideh2015}, \citealt{karamehmetoglu2019}) show strong emission lines of helium, suggesting recent mass loss from stripped progenitors. In both Types IIn and Ibn, there is evidence that in at least some cases, the mass-loss is generated by precursor events, prior to the SN explosion (e.g. \citealt{pastorello2007}, \citealt{foley2007}, \citealt{ofek2014}, Strotjohann et al., in prep.)

If the extension of the CSM is confined to a relatively compact location around an exploding star or if its density is low, the explosion shock-breakout flash may ionize the CSM. The resulting recombination  emission lines will be transient, persisting only until the SN ejecta overtake and engulf the denser parts of the CSM (supernovae with``flash ionized'' emission lines; \citealt{galyam2014}). Such events later evolve spectroscopically in a regular manner, e.g., presenting photospheric spectra with broad P-Cygni line profiles. 

Several serendipitous observation of such ``flash features'' in early supernova spectra were made over the years (e.g., \citealt{niemela1985}, \citealt{garnavich1994}, \citealt{quimby2007}). We define flash features here as transient narrow emission lines (of the order of $\approx 10^2$ km s$^{-1}$) of highly ionised species (e.g.: \ion{He}{2}, \ion{C}{3}, \ion{N}{3}, \ion{N}{4}) in the early phases of the supernova event (less than a week, in general from estimated explosion). \cite{galyam2014} presented very early observations of the Type IIb SN 2013cu, and noted that such flash features could be routinely observed by modern high-cadence SN surveys and probe the composition of the pre-explosion mass loss, and hence the surface composition of the progenitor star, which is hard to measure by other means. This motivated additional work on such flash objects. For example, \cite{yaron2017} presented a time-series of early spectra and used it to constrain the CSM distribution around the spectroscopically normal SN 2013fs, showing that the CSM was lost from the progenitor in the year prior to its explosion. 
\cite{Hosseinzadeh2018} studied the low-luminosity Type II event SN~2016bkv which showed early flash ionisation features. They suggest that its early light-curve bump could suggest a contribution from CSM interaction to the early light curve, motivating the systematic study of early light curves of Type II SNe showing flash features to distinguish between properties originating from the CSM (e.g., perhaps, peak luminosity) and those determined by the progenitors via shock cooling emission. 
Several theoretical investigations also focused on such events (e.g., \citealt{groh2014}, \citealt{dessart2017}, \citealt{moriya2017} and \citealt{grohb2020}). 

A systematic study of such transient signatures of CSM around SN II progenitor stars has been limited by the challenge of routinely observing CC SNe early enough (typically within less than a few days from explosion), before these features disappear. \cite{khazov2016} conducted the first study of the occurrence of flash ionisation in Type II SNe using data from the PTF and iPTF surveys, and gathered 12 objects showing flash ionisation features. They estimate that more than $\sim20\%$ of SNe II show flash ionisation features, but their analysis is limited by the heterogeneity of their data. 

Routine and systematic observations of young (``infant'') SNe was one of the main goals of the ZTF survey \citep{galyam2019,grahamztf2019}. Here, we present our systematic search  and follow-up observations of infant Type II SNe from ZTF. We use a sample of 28 events collected during the first year of ZTF operation, ten of which were spectroscopically observed within two days of estimated explosion, to place a lower limit on the fraction of SN progenitor stars embedded in CSM.  

In section $\S~2$, we describe the properties of our infant SN survey and the construction of our sample of SNe II. In $\S~3$ we present our analysis, in $\S~4$ we discuss our findings, and we conclude in $\S~5$.

\vspace{2.5cm}

\section{Observations and Sample construction}
\subsection{Selecting infant SNe from the ZTF partnership stream}
The Zwicky Transient Facility (ZTF) is a wide-field, high cadence, multiband survey that started operating in March 2018 \citep{bellmztf2019,grahamztf2019}. ZTF imaging is obtained using the Samuel Oschin 48" Schmidt telescope at Palomar observatory (P48). ZTF observing time is divided among three programs: the public (MSIP) 3-day all-sky survey, partnership surveys, and Caltech programs. This paper is based on data obtained by the high-cadence partnership survey. As part of this program, during 2018, extra-galactic survey fields were observed in both the ZTF g- and r-bands 2-3 times per night per band. New images were processed through the ZTF pipeline \citep{masciztf2019} and reference images built by combining stacks of previous ZTF imaging in each band were then subtracted using the \cite{ZOGY2016} image subtraction algorithm (ZOGY).  A 30s integration time was used in both $g-$ and $r-$band exposures, and a 5$\sigma$ detection limit down to $\sim$20.5 mag in $r$ can be reached in a single observation. 

We conducted our year-1 ZTF survey for infant SNe following the methodology of \cite{galyam2011}. We selected potential targets via a custom filter running on the ZTF alert stream using the GROWTH Marshal platform \citep{Kasliwal2019}. 
The filter scheme was based on the criteria listed in Table~\ref{filter-table}.

\begin{table*}
    \centering
    \caption{Filter criteria selecting infant SN candidates}
    \begin{tabular}{|l|l|}
        \hline
         \textit{Stationary} & Reject solar-system objects using apparent motion \\
         \hline
         \textit{Recent limit} & Require a non-detection limit within $< 2.5$ days from the first detection\\
         \hline
         \textit{Extragalactic} & Reject alerts within 14 degrees from the Galactic plane \\
         \hline
         \textit{Significant} & Require a ZOGY score of $> 5$\\
         \hline
         \textit{Stellar} & Require a SG (star-galaxy) score of $> 0.49$\\
         \hline
    \end{tabular}
    \label{filter-table}
\end{table*}

Alerts that passed our filter (typically $50-100$ alerts per day) were then visually scanned by a duty astronomer, in order to reject various artefacts (such as unmasked bad pixels or ghosts) and false positive signals, such as flaring M stars, CVs and AGN. Most spurious sources could be identified by cross matching with additional catalogues (e.g., \textit{WISE} IR photometry \citep{wise2010} to detect red M stars, the Gaia DR2 catalog \citep{gaiadr22018}, and catalogs from time domain surveys such as the Palomar Transient Factory (PTF; \citealt{law2009}) and the Catalina Real-Time Survey (CRTS; \citealt{drake2014}) for previous variability of CVs and AGN. 

Due to time-zone differences, our scanning team (located mostly at the Weizmann Institute in Israel and at the Oskar Klein Center (OKC) in Sweden) could routinely scan the incoming alert stream during the California night time, with the goal of triggering spectroscopic follow-up of promising infant SN candidates within hours of discovery (and thus typically within $<2$\,days from explosion), as well as {\it Swift} \citep{swift2004} Target-of-Opportunity (ToO) UV photometry.\\

\begin{figure}
    \centering
    \includegraphics[scale=0.35]{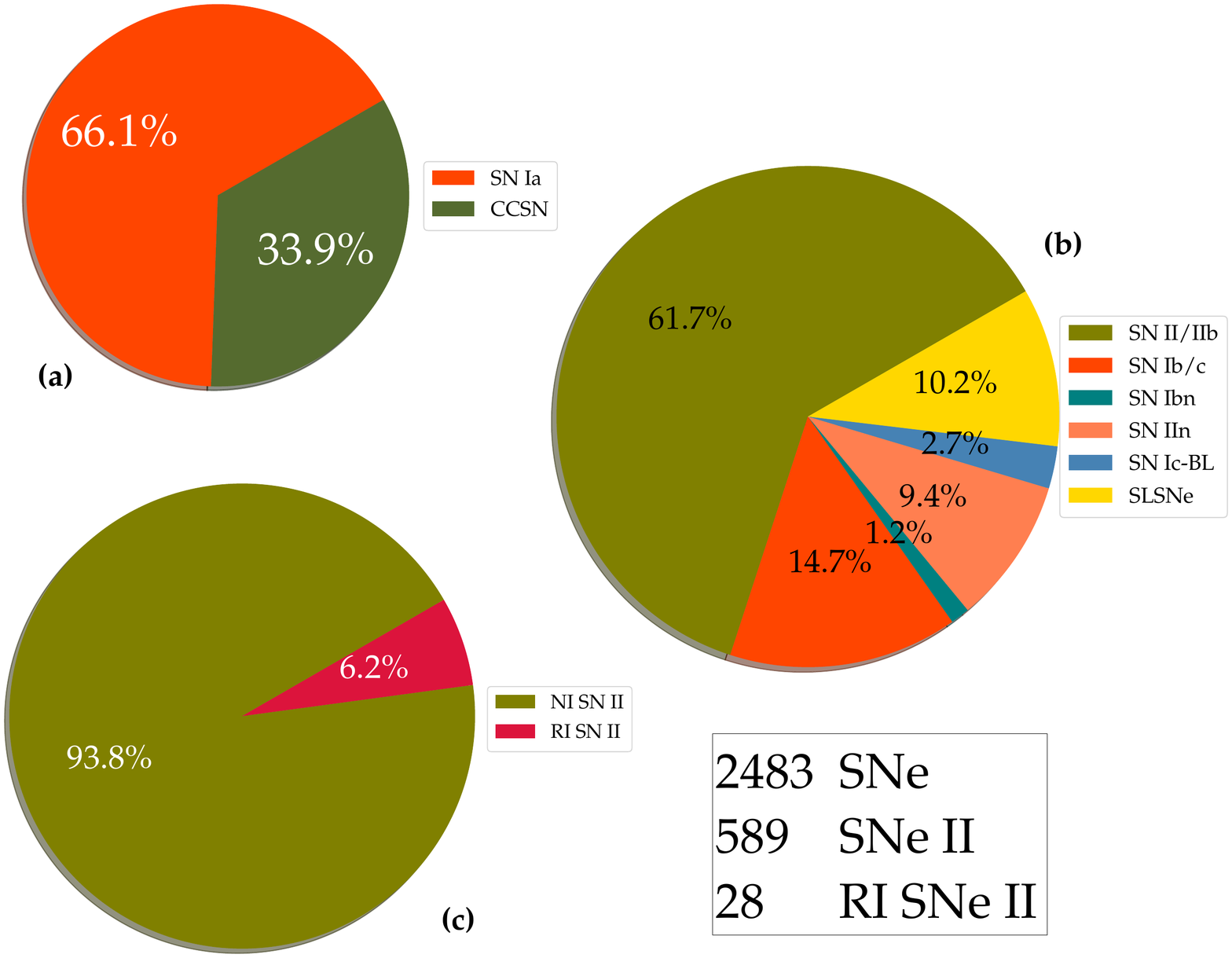}
    \caption{ZTF Spectroscopically-confirmed SN discovery statistics during 2018. (a) Most events ($63\%$) are SNe Ia; CC SNe comprise about $37\%$. (b) The division among CC SN sub-classes (c) The fraction of real infant (RI) SNe II is $4.8\%$ of the total Type II population. NI stands for the Non Infant SN II population (see text).  }
    \label{fig:SN_stats}
\end{figure}

\subsection{Sample Construction}
\label{sec:sample}

Figure ~\ref{fig:SN_stats} shows the SN Type distribution amongst the $\sim2500$ spectroscopically-confirmed SNe gathered by ZTF between March and December 2018. About 37\% are core-collapse events, and $\sim63\%$ of those are of Type II. Since the large majority of flash events are SNe II, we can only place statistically meaningful constraints on the frequency of this phenomenon among Type II SNe. We therefore analyze here this population only. 

Our infant SN program allowed us to obtain early photometric and spectroscopic follow-up of young SNe. However, it is possible that we have missed some relevant candidates. In order to ensure the completeness of our sample, we therefore inspected all spectroscopically classified SNe II (including subtypes IIn and IIb) from ZTF \footnote{between March 2018 and December 2018} using the \texttt{ZTFquery} package \citep{mickael_rigault_2018_1345222}. We pulled from this sample all events (the large majority) lacking a ZTF non-detection limit within 2.5 days prior to the first detection recorded on the ZTF Marshal. To include events in our final sample, we required that they show significant and rapid increase in flux, as previously observed for very young SNe (e.g., \citealt{galyam2014}, \citealt{yaron2017}), with respect to the last non-detection. This excludes older events that are just slightly below our detection limit and are picked up by the filter when they slowly rise, or when conditions improve. We implemented a cut on the observed rise of $\Delta$r or $\Delta$g$>0.5$\,mag with respect to the recent limit in the same band, and note all events that satisfy this cut as ``real infant'' (RI; Fig.~\ref{fig:SN_stats}, panel C).  

All in all, we gathered 43 candidates which fulfilled the RI criteria. Additional inspection led us to determine that 15 candidates were spurious (see Appendix A for details). Our final sample (Table~\ref{Master_table}) thus includes a total of 28 RI Type II SNe, or about 5\% of all the SNe II found by the ZTF survey for 2018. During its first year of operation (starting March 2018), ZTF obtained useful observations for our program during approximately 32 weeks, excluding periods of reference image building (initially), periods dedicated to Galactic observations, and periods of technical/weather closure. We therefore find that the survey provided about one real infant SN II per week.    

\subsection{Spectroscopic Observations}

Our goal was to obtain rapid spectroscopy of RI SN candidates following the methods of \cite{galyam2011}. This was made possible using 
rapid ToO follow-up programs as well as on-request access to scheduled nights on various telescopes. During the scanning campaign, we applied the following criteria for rapid spectroscopic triggers. The robotic SEDm (see below) was triggered for all candidates brighter than a threshold limiting magnitude ($19$\,mag during 2018). Higher resolution spectra (using WHT, Gemini or other available instruments) were triggered for events showing recent non-detection limits (within $2.5$\,d prior to first detection) as well as a significant rise in magnitude compared to a recent limit or within the observing night. 

\paragraph{P60/SEDm} The Spectral Energy Distribution Machine (SEDm; \citealt{benami2012,blagorodnova2018,neillSEDm2019}) is a high-throughput, low-resolution spectrograph, mounted on the 60" robotic telescope (P60; \citealt{cenko2006}) at Palomar observatory. $65\%$ of the time on the SEDm was  dedicated to ZTF partnership follow up. SEDm data are reduced using an automated pipeline (\citealt{rigaultsedm2019}). The co-location of the P60 and ZTF/P48 on the same mountain, as well as the P60 robotic response capability, enable very short (often same-night) response to ZTF events, sometimes very close to the time of first detection (e.g., see ZTF18abwlsoi, below). However, the low resolution ($R\sim100$) of the instrument limits our capability to characterise narrow emission lines. This, along with the overall sensitivity of the system, motivated us to try to obtain higher-resolution follow-up spectroscopy with other, larger, telescopes, in particular for all infant SNe detected below a magnitude cut of $r\sim19$ mag.  

\paragraph{P200/DBSP} We used the Double Beam SPectrograph (DBSP; \citealt{DBSP1982}) mounted on the 5m Hale telescope at Palomar Observatory (P200) to obtain follow-up spectroscopy in either ToO mode or during  classically scheduled nights. The default configuration used the 600/4000 grism on the blue side, the 316/7150 grating on the red side, along with the D55 dichroic, achieving a spectral resolution $R\sim 1000$. Spectra obtained with DBSP were reduced using the pyraf-dbsp pipeline \citep{bellm2016}. 

\paragraph{WHT-ISIS/ACAM} We obtained access to the 4.2m William Herschel Telescope (WHT) at the Observatorio del Roque de los Muchachos in La Palma, Spain via the Optical Infrared Coordination Network for Astronomy (OPTICON\footnote{https://www.astro-opticon.org/index.html}) program\footnote{Program IDs OPT/2017B/053, OPT/2018B/011, OPT/2019A/024, PI Gal-Yam}. We used both single-slit spectrographs ISIS and ACAM \citep{ACAM2008} in ToO service observing mode. The delivered resolutions were $R\sim1000$ and $R\sim400$, respectively. Spectral data were reduced using standard routines within IRAF\footnote{{IRAF} is distributed by the National Optical Astronomy Observatories, which are operated by the Association of Universities for Research in Astronomy, Inc., under cooperative agreement with the National Science Foundation.}.

\paragraph{Keck/LRIS} We used the Low-Resolution Imaging Spectrometer (LRIS; \citealt{Keck1995}) mounted on the Keck-I 10m telecope at the W. M. Keck Observatory in Hawaii in either ToO mode or during scheduled nights. The data were reduced using the LRIS automated reduction pipeline Lpipe \citep{perley2019}.

\paragraph{GMOS/Gemini} We used the Gemini Multi-Object Spectrographs (GMOS; \citealt{Hook2004}) mounted on the Gemini North 8m telescope at the Gemini Observatory on Mauna Kea, Hawaii. All observations were conducted at small airmass ($\lesssim1.2$). For each SN, we obtained 2$\times$900\,s exposures using the B600 grating with central wavelengths of $520$\,nm and $525$\,nm. The $5$\,nm shift in the effective central wavelength was applied to cover the chip gap, yielding a total integration time of $3600$\,s. A 1.0$\arcsec$-wide slit was placed on each target at the parallactic angle. The GMOS data were reduced following standard procedures using the Gemini IRAF package.  

\paragraph{APO/DIS} We used the Dual Imaging Spectrograph (DIS) on the Astrophysical Research Consortium (ARC) 3.5\,m telescope at Apache Point Observatory (APO) during scheduled nights. The data were reduced using standard procedures and calibrated to a standard star obtained on the same night using the PyDIS package \citep{davenport}.


All the data presented in this paper will be made public on WISeREP \citep{wiserep2012}. \\

\begin{table*}[ht]
\caption{Sample of Real Infant 2018 (28 objects) }
\begin{tabular}{lllllllllll}
IAU & Internal & Type \footnote{Classification reports referenced in square brackets} & Redshift & Explosion & Error & First & Last non & First   & Telescope/ & Flash \\
name & ZTF & & z & JD Date &  & detection & detection & spectrum & instrument & \\   
(SN) & name & & & [d] & [d] & [d] \footnote{All times reported relative to the estimated explosion date in fractional days} & [d] & [d] & & \\
\hline
\hline
2018grf & 18abwlsoi & SN II [\citenum{ZTF18abwlsoi}]& 0.050 & 2458377.6103 & 0.0139 & 0.0227 & -0.8725 & 0.1407 & P60/SEDm & \cmark \\
2018fzn & 18abojpnr & SN IIb [\citenum{ZTF18abojpnr}] & 0.037 & 2458351.7068 & 0.0103 & 0.0102 & -0.0103 & 0.1902 & P60/SEDm & \xmark \\
2018dfi & 18abffyqp & SN IIb [\citenum{ZTF18abffyqp}] & 0.031 & 2458307.2540 & 0.4320 & 0.4320 & -0.4320 & 0.6180 & P200/DBSP & \cmark \\
2018cxn & 18abckutn & SN II [\citenum{ZTF18abckutn}] & 0.040 & 2458289.8074 & 0.4189 & 0.0576 & -0.0494 & 0.9406 & P200/DBSP & \xmark \\
2018dfc & 18abeajml & SN II [\citenum{ZTF18abeajml}] & 0.037 & 2458303.7777 & 0.0118 & 0.0213 & -0.9806 & 1.0153 & P60/SEDm & \cmark \\
2018fif & 18abokyfk & SN II [\citenum{ZTF18abokyfk}] & 0.017 & 2458350.9535 & 0.3743 & -0.0635 & -1.0525 & 1.0525 & P200/DBSP & \cmark \\
2018gts & 18abvvmdf & SN II [\citenum{ZTF18abvvmdf}] & 0.030 & 2458375.1028 & 0.5551 & -0.4688 & -1.3648 & 1.5162 & P60/SEDm & \cmark \\
2018cyg & 18abdbysy & SN II [\citenum{ZTF18abdbysy}] & 0.011 & 2458294.7273 & 0.2034 & 0.0297 & 0.0147 & 1.6727 & WHT/ACAM & \qmark \\
2018cug & 18abcptmt & SN II [\citenum{ZTF18abcptmt}] & 0.050 & 2458290.9160 & 0.0250 & -0.0066 & -0.0670 & 1.7960 & P60/SEDm & \cmark \\
2018egh & 18abgqvwv & SN II [\citenum{classrep_rb}] & 0.038 & 2458312.7454 & 0.4351 & 0.9846 & 0.0931 & 1.8236 & WHT/ISIS & \qmark \\
\hline
2018bqs & 18aarpttw & SN II [\citenum{classrep_rb}] & 0.047 & 2458246.8133 & 0.0071 & 0.0087 & -0.9926 & 2.0867 & APO/DIS & \xmark \\
2018fsm & 18absldfl & SN II [\citenum{ZTF18absldfl}] & 0.040 & 2458363.4226 & 0.4565 & 0.4564 & -0.4564 & 2.3674 & P60/SEDm & \xmark \\
2018bge & 18aaqkoyr & SN II [\citenum{ZTF18aaqkoyr}] & 0.023 & 2458243.1671 & 0.5180 & 0.5179 & -0.5180 & 2.5169 & P200/DBSP & \xmark \\
2018leh & 18adbmrug & SN IIn [\citenum{ZTF18adbmrug}] & 0.024 & 2458481.7505 & 0.9485 & 0.9485 & -0.9485 & 3.6985 & WHT/ISIS & \cmark \\
2018iua & 18acploez & SN II [\citenum{classrep_rb}] & 0.040 & 2458439.9877 & 0.9784 & 0.9783 & -0.9783 & 3.7933 & P60/SEDm & \xmark \\
2018gvn & 18abyvenk & SN II [\citenum{ZTF18abyvenk}] & 0.040 & 2458385.6198 & 0.0011 & 0.0012 & -0.8565 & 6.1122 & P60/SEDm & \xmark \\
2018clq & 18aatlfus & SN II [\citenum{ZTF18aatlfus}] & 0.045 & 2458248.8967 & 0.9564 & 0.9564 & -0.9564 & 6.9274 & P60/SEDm & \xmark \\
\hline
2018ccp & 18aawyjjq & SN II [\citenum{ZTF18aawyjjq}] & 0.040 & 2458263.7743 & 0.1241 & 0.0106 & -0.8684 & 8.1087 & P60/SEDm & \xmark \\
2018lth & 18aayxxew & SN II [\citenum{classrep_rb}] & 0.061 & 2458278.6531 & 0.9154 & 0.0509 & -1.9102 & 8.1589 & Keck/LRIS & \xmark \\
2018inm & 18achtnvk & SN II [\citenum{ZTF18achtnvk}] & 0.040 & 2458432.9113 & 0.6895 & 1.9927 & 1.9497 & 9.0137 & P60/SEDm & \xmark \\
2018iwe & 18abufaej & SN II [\citenum{classrep_rb}] & 0.062 & 2458368.8561 & 0.0179 & 0.0179 & -0.0179 & 12.0159 & P60/SEDm & \xmark \\
2018fso & 18abrlljc & SN II [\citenum{ZTF18abrlljc}] & 0.050 & 2458357.6987 & 0.8255 & -0.0177 & -0.9157 & 14.0113 & P60/SEDm & \xmark \\
2018efd & 18abgrbjb & SN IIb [\citenum{ZTF18abgrbjb}] & 0.030 & 2458312.8922 & 0.3938 & 0.8568 & 0.8244 & 14.9388 & P60/SEDm & \xmark \\
2018cyh & 18abcezmh & SN II [\citenum{classrep_rb}] & 0.057 & 2458286.3752 & 0.6050 & 0.4348 & 0.3898 & 16.5678 & P60/SEDm & \xmark \\
2018ltg & 18aarqxbw & SN II [\citenum{classrep_rb}] & 0.048 & 2458241.4360 & 3.4950 & 3.4950 & -3.4950 & 37.5310 & P200/DBSP & \xmark \\
2018lti & 18abddjpt & SN II [\citenum{classrep_rb}] & 0.070 & 2458294.6217 & 0.1224 & 0.1693 & -0.7917 & 40.2333 & P60/SEDm & \xmark \\
2018efj & 18abimhfu & SN II [\citenum{ZTF18abimhfu}] & 0.050 & 2458320.6574 & 0.0210 & 0.0096 & -0.9028 & 42.0096 & P60/SEDm & \xmark \\
2018cfj & 18aavpady & SN II [\citenum{classrep_rb}] & 0.047 & 2458256.4531 & 0.4771 & 0.4771 & -0.4771 & 55.0469 & Keck/LRIS & \xmark \\
\hline
\end{tabular}
\label{Master_table}
\end{table*}

\subsection{Photometry}
The ZTF alert system \citep{Patterson_2018} provides on the fly photometry \citep{masciztf2019} and astrometry based on a single image for each alert.  In order to improve our photometric measurements (and in particular, to test the validity of non-detections just prior to discovery) we performed forced PSF photometry at the location of each event. As shown by \cite{yaron2019}, the $95\%$ astrometric scatter among ZTF alerts is $\sim0.44$''; for our events we have multiple detections, with typically higher signal-to-noise ratio data around the SN peak compared to the initial first detections. We therefore compute the median coordinates of all the alert packages and perform forced photometry using this improved astrometric location. 

We use the pipeline developed by F. Masci and R. Laher\footnote{http://web.ipac.caltech.edu/staff/fmasci/ztf/forcedphot.pdf} to perform forced PSF photometry at the median SN centroid on the ZTF difference images available from the IRSA database . For each light curve, we filter out measurements returned by the pipeline with non-valid flux values.  

We perform a further quality cut on each light curve by rejecting observations with a data quality parameter \textit{scisigpix}\footnote{A parameter calculated by the pipeline that measures the pixel noise in each science image} that is more than 5 times the median absolute deviation (MAD) away from the median of this parameter for each light curve. We also remove faulty measurements where the \textit{infobitssci} parameter is not zero.  
According to the Masci \& Laher prescription we rescale the flux errors by the square root of the $\chi^2$ of the PSF fit estimate in each image. We then correct each measured forced photometry flux value by the photometric zero point of each image, as provided by the pipeline:

\begin{equation}
    f_{zp,corrected} = f_{forced-phot} \times  10^{-0.4 \times z_p}
\end{equation}

We determine our zero-flux baseline using forced photometry observations obtained prior to the SN explosion. We calculate the median of these observations, reject outliers that are $>3$\,MAD away from the median, re-calculate the median and subtract it from our measured post-explosion flux values; these corrections are typically very small, of the order of $<0.1\%$ of the supernova flux values.

If the ratio between the measured flux and the uncertainty $\sigma$ is below $3$, we consider this measurement as a non-detection, and report a 5$\sigma$ upper limit. Otherwise (if the flux to error ratio is above $3\sigma$) we report the flux, magnitude and respective errors. Finally, we correct for Galactic extinction using the python package \textit{extinction}\footnote{https://github.com/kbarbary/extinction}, using local extinction values from \cite{schlafly2011} and assuming a selective extinction of $R_V = 3.1$ and the \cite{cardelli1989} extinction law. We also correct the light curves to restframe time according to the spectroscopic redshifts. 

We recovered detections prior to the first detection by the real-time pipeline using the forced photometry pipeline in 11 cases \footnote{ZTF18aarqxbw, ZTF18aavpady, ZTF18aawyjjq, ZTF18abcezmh, ZTF18abckutn, ZTF18abcptmt, ZTF18abdbysy, ZTF18abddjpt, ZTF18abokyfk, ZTF18abrlljc, ZTF18abvvmdf}. We redefined the first detection and last non-detection according to the forced photometry pipeline measurements in these cases. 

We present our photometry for all RI objects in Table~\ref{phot-table}. 

\begin{table*}
    \centering
    \caption{Forced photometry of the RI sample}
    \begin{tabular}{lccccccc}
    Object & Filter & JD & Flux & Flux error & Apparent magnitude & Absolute magnitude & Magnitude error  \\
    &  &  & [$10^{-8}$\, Mgy] & [$10^{-8}$\,Mgy] & [AB mag] & [AB mag] & [AB mag] \\
    \hline
    \hline
    ... & ... & ... & ... & ... & ... & ... & ...  \\
    ZTF18aarpttw & g & 2458258.8522 & 2.3555 & 0.0868 & 19.07 & -17.60 & 0.04 \\
    ZTF18aatlfus & g & 2458258.8564 & 3.9348 & 0.0916 & 18.51 & -18.06 & 0.03 \\
    ZTF18aarqxbw & g & 2458258.8624 & 1.4655 & 0.0709 & 19.59 & -17.13 & 0.05 \\
    ZTF18aarqxbw & g & 2458258.8634 & 1.4371 & 0.0731 & 19.61 & -17.11 & 0.06 \\
    ZTF18aavpady & g & 2458258.8672 & 1.2260 & 0.0633 & 19.78 & -16.89 & 0.06 \\
    ... & ... & ... & ... & ... & ... & ... & ...  \\
    \hline
    \end{tabular}
    \tablecomments{This table includes the flux measurements returned by the forced photometry pipeline. In this table, we report the last non detections within 2.5 days from the first marshal detection and all the measurements which follow. The full version of this table is electronic.}
    \label{phot-table}
\end{table*}

\section{Analysis and Results}
In this section, we study the 28 RI SNe that passed our selection criteria, excluding spurious candidates (see Appendix A for details). In order to measure the fraction of objects showing flash features and thus evidence for CSM, we estimate the explosion time based on ZTF forced photometry light curves. We then define subsamples based on the SN age (relative to estimated explosion) at the time the first spectrum was obtained. 

\subsection{Explosion time estimation}
\begin{figure}
\hspace{-1.25cm}
    \includegraphics[scale=0.35]{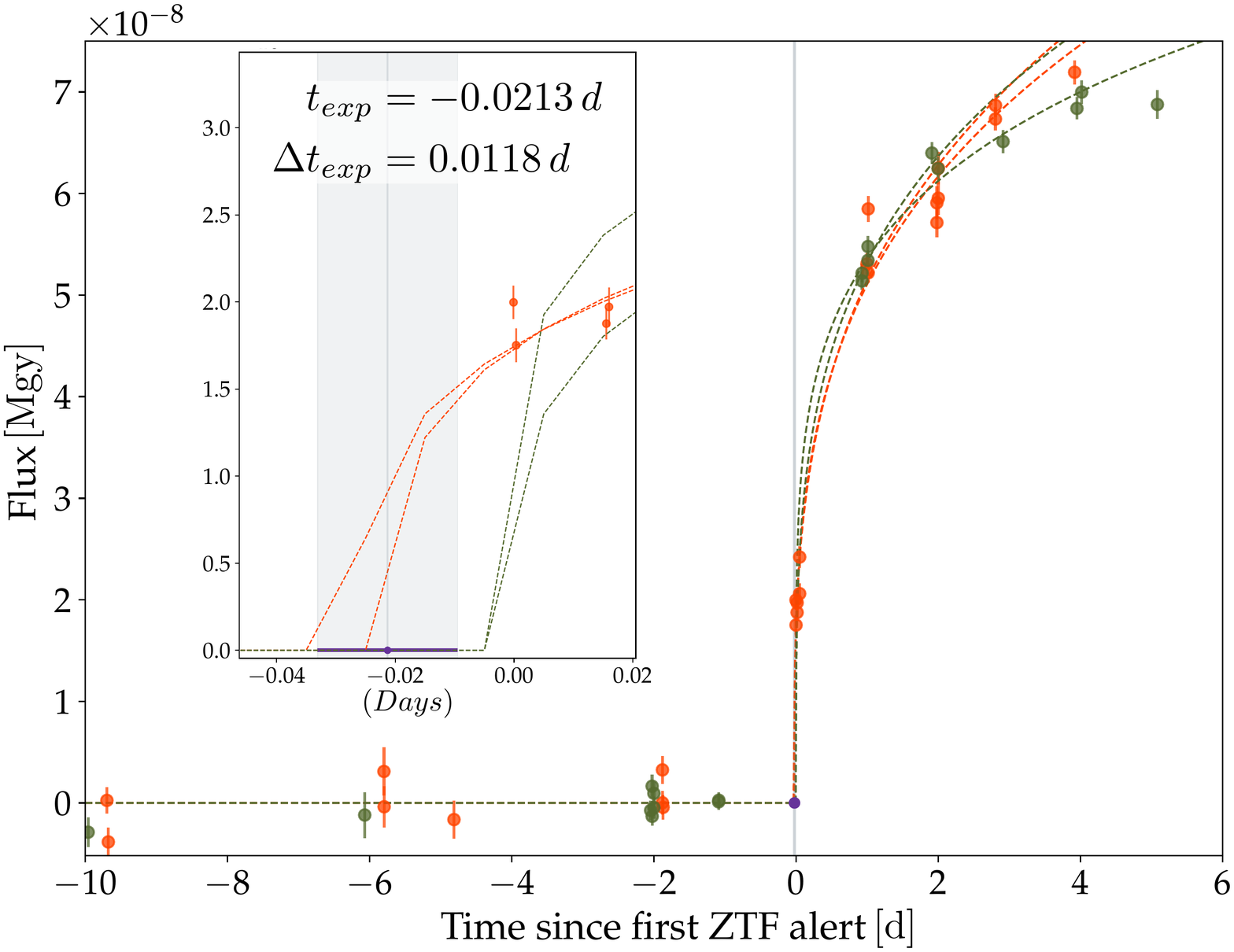}
    \caption{Early light curve fits used to determine the explosion date for SN 2018dfc. Power-law fits to the observations during the first 2 or 5 days are shown in both the $g$ (green points) and $r$ (red points) bands. The mean and standard deviation of the fits (inset) are adopted as the the explosion time and the error. The time origin is defined as the time of the first alert from ZTF. }
    \label{fig:LC}
\end{figure}

In order to estimate the explosion time, which we define here as the time of zero-flux, we fit a general power law of the form to our flux measurements: 
\begin{equation}
    f(t)  = a \times (t-t_{exp}) ^ n 
\end{equation}
using the routine \texttt{curvefit} within the {\it astropy} python package (\citealt{astropy2013}). We fit the first 2 days of data following the first detection as well as the first 5 days (see Fig.~\ref{fig:LC} for example) in both the $g$ and $r-$bands.   
The estimated explosion time is taken as the weighted mean of the four fits, and we adopt the standard deviation as the error on this value. In ten cases, however, there were not enough data in either band to perform the fit. In those cases, we set the explosion date as the mean between the time of the last non detection and the first detection. In all but four of the cases the estimated explosion date is within less than a day from the first detection (Fig.~\ref{fig:timeline}; Table~\ref{Master_table}).

\subsection{Peak magnitude}

\begin{figure}
\hspace{-0.5cm}
    \includegraphics[scale=0.45]{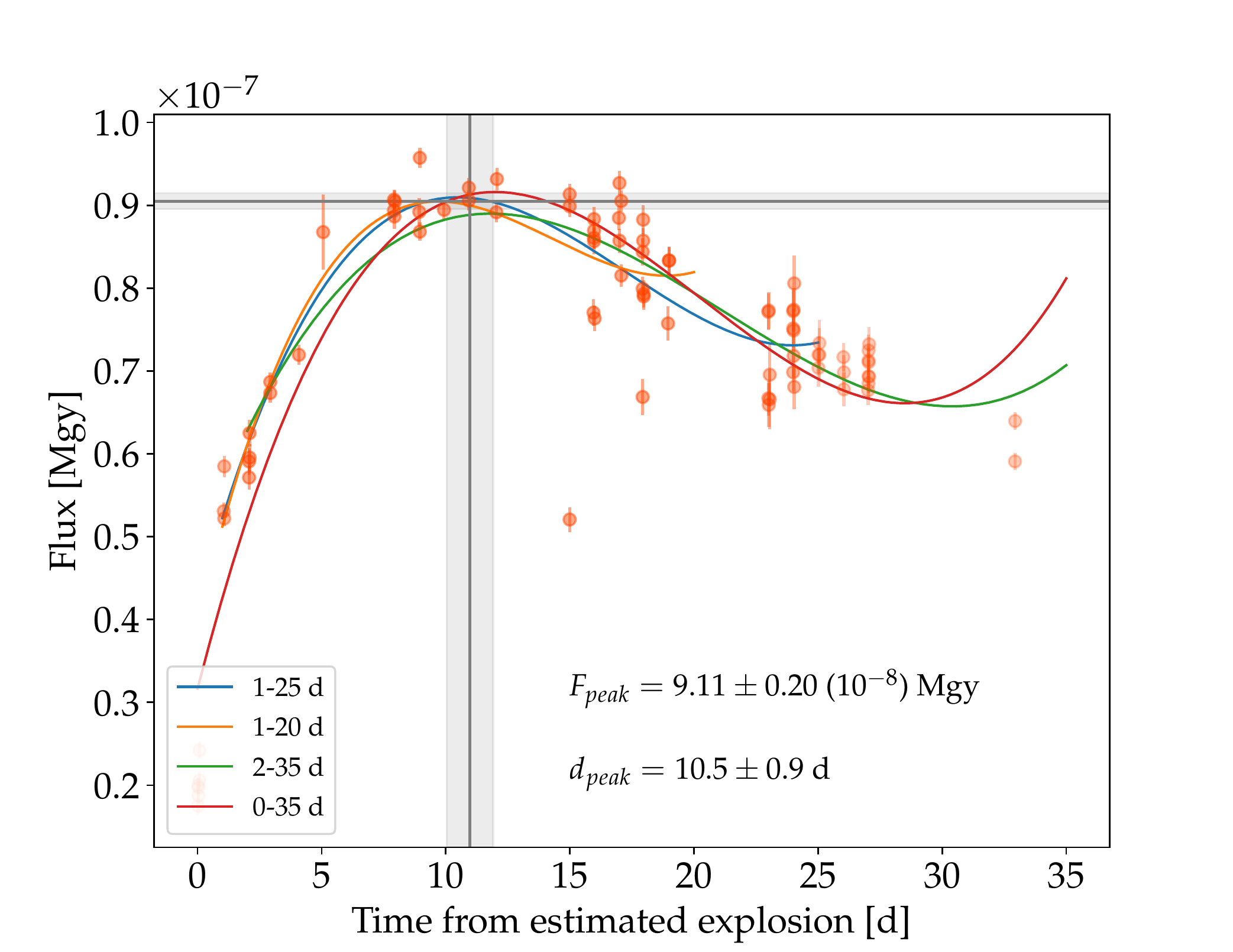}
    \caption{Example of the peak estimation in the red band for SN2018dfc. The different curves correspond to a polynomial of order 3 fitted over the time intervals noted in the legend. The cross corresponds to the peak date and flux estimated from the mean of all the values obtained, and the grey bands note the estimated errors, see text for details.}
    \label{fig:peak_mag}
\end{figure}

Following \cite{khazov2016}, we also test if events showing flash features are on average more luminous. As can be seen from Table~\ref{Master_table}, the relevant events to consider are only those with relatively early spectra. We therefore compute the peak magnitude of all seventeen events with a first spectrum obtained within $7$\,days from explosion. 
We use the forced photometry lightcurves to evaluate the peak magnitude. We fit a polynomial of order 3 to the flux measurements, over several intervals of time whose lower bound is within the first few days from explosion time and upper bound between 10 to 40 days after the estimated explosion time (Fig.~\ref{fig:peak_mag}). We adopt the mean and range of peak times obtained from these fits as the peak date and its error (vertical grey band in Fig.~\ref{fig:peak_mag}) and take the mean and standard deviation of the flux value within this range to be the peak flux and error (horizontal grey band in Fig.~\ref{fig:peak_mag}). We report these values for each event in each band in Table~\ref{mag_table}.

\subsection{Early spectroscopy}

\begin{figure}
    \vspace{1cm}
    \includegraphics[trim = 100 50 100 100 ,scale = 0.35]{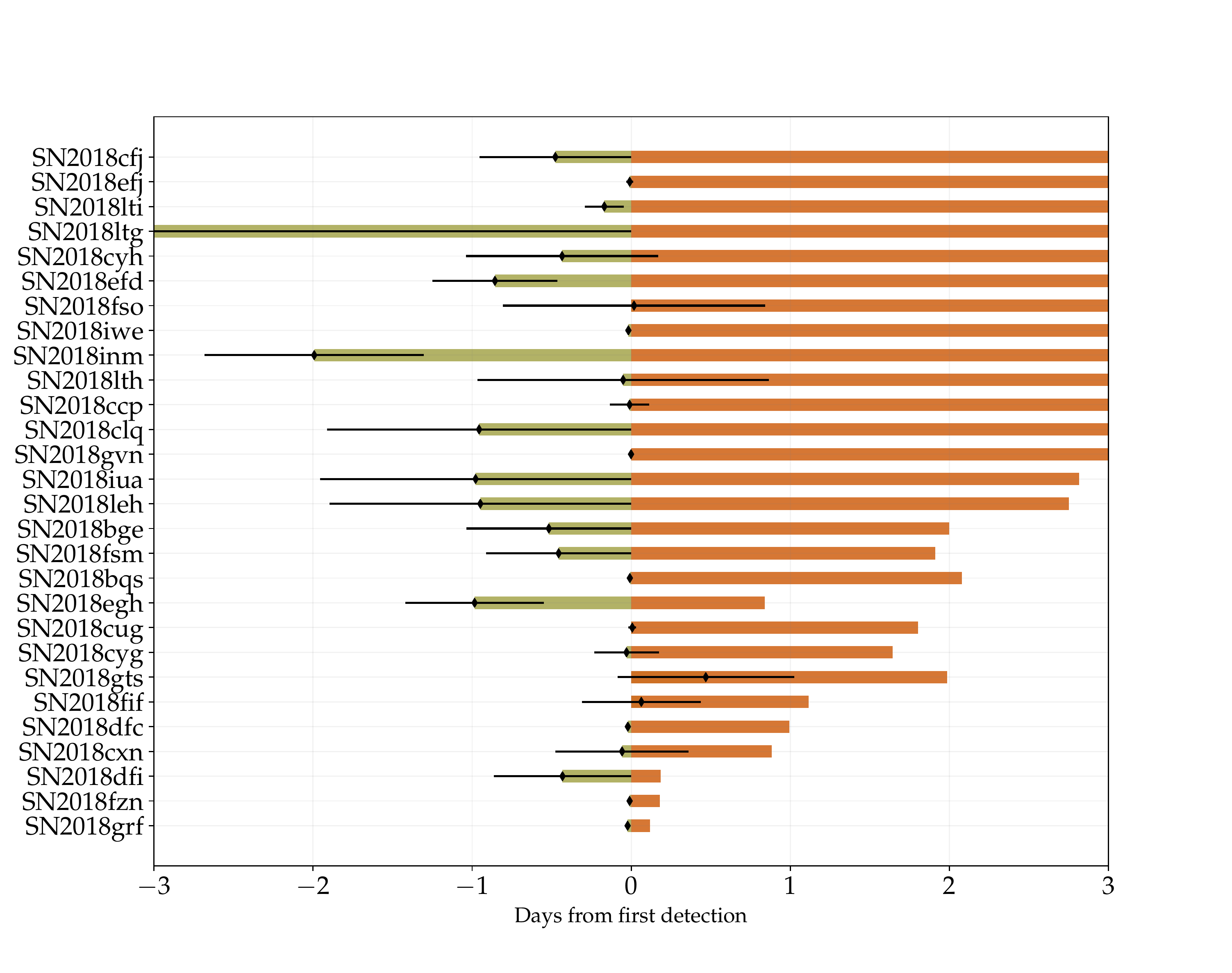}
    \caption{A graphic summary of the sample timeline, from the estimated explosion date (green) to the time of the first spectrum (red). The x-axis origin (``0'' time) corresponds to the first photometric detection of each candidate. SN 2018ltg was still included in the sample of RI SNe II since its non-detection limit in the Marshal alert system was $<2.5\,d$ although the explosion time estimation with the forced photometry lightcurve puts the limits to more than three days.}
    \label{fig:timeline}
\end{figure}

We sort the 28 RI SNe in our sample according to the difference between the estimated explosion time and the time of first spectrum (Table~\ref{Master_table}, ``First spectrum'' column; Fig.~\ref{fig:timeline}).
 From previous work (\citealt{galyam2014}, \citealt{yaron2017}, \citealt{khazov2016}), we know that flash features are typically present from the time of explosion up to several days later. We therefore define a sub-sample including events with spectra obtained within 2\,d from explosion (top of Table~\ref{Master_table}). For about one third of the total sample (ten objects) we have been able to secure a first spectrum within less than 2 days from the estimated explosion time.

Throughout the 2018 campaign, we find that seven infant supernovae of Type II show flash features (Table~\ref{Master_table}; Fig~\ref{fig:flashers}). Two additional infant objects were marked as potential flash events (Fig.~\ref{fig:bumpers}; see below). Four of the seven confirmed flashers had their first spectrum obtained with SEDm. 

The two-day sub-sample we are considering includes 6 events showing flash features (one object, SN 2018leh, shows flash features but its first spectrum was obtained only $>3$\,days after explosion, Table~\ref{Master_table}), the two potential flashers, and  two events have high signal to noise early spectra that show no flash features (Fig.~\ref{fig:no_flashers}). 

\begin{table*}
\centering
\caption{Peak absolute magnitudes of the 17 objects within the 7-day spectroscopic sub-sample}
\begin{tabular}{lllllllll}
IAU & Redshift & Distance & Filter & Rise time  & Peak Flux  & Peak & Peak & Magnitude \\
name & z & modulus & & to peak\footnote{From estimated explosion time} & $10^{-8}$[Jy] & apparent & absolute & error \\
 & & [mag] & & [d] & & AB mag & AB mag & \\
\hline
\hline
SN2018bge & 0.023 & 35.08 & g & 9.9 $\pm$ 0.6 & 6.94 $\pm$ 0.10 & 17.90 & -17.18 & 0.02 \\
 &  &  & r & 18.1 $\pm$ 1.4 & 7.45 $\pm$ 0.05 & 17.82 & -17.26 & 0.01 \\
SN2018bqs & 0.047 & 36.67 & g & 6.1 $\pm$ 0.3 & 3.25 $\pm$ 0.10 & 18.72 & -17.95 & 0.03 \\
 &  &  & r & 8.6 $\pm$ 0.7 & 3.20 $\pm$ 0.03 & 18.74 & -17.93 & 0.01 \\
SN2018clq & 0.045 & 36.57 & g & 4.7 $\pm$ 1.2 & 5.93 $\pm$ 0.10 & 18.07 & -18.51 & 0.02 \\
 &  &  & r & 5.6 $\pm$ 1.5 & 5.51 $\pm$ 0.50 & 18.15 & -18.43 & 0.10 \\
SN2018cxn & 0.040 & 36.31 & g & 9.8 $\pm$ 1.3 & 2.90 $\pm$ 0.03 & 18.84 & -17.47 & 0.01 \\
 &  &  & r & 15.8 $\pm$ 0.7 & 2.86 $\pm$ 0.02 & 18.86 & -17.45 & 0.01 \\
SN2018cug & 0.050 & 36.81 & g & 8.0 $\pm$ 1.2 & 3.82 $\pm$ 0.06 & 18.54 & -18.26 & 0.02 \\
 &  &  & r & 10.3 $\pm$ 0.9 & 3.72 $\pm$ 0.05 & 18.57 & -18.23 & 0.01 \\
SN2018cyg & 0.011 & 33.51 & g & 10.8 $\pm$ 0.6 & 2.14 $\pm$ 0.03 & 19.17 & -14.33 & 0.02 \\
 &  &  & r & 16.3 $\pm$ 0.9 & 5.37 $\pm$ 0.20 & 18.18 & -15.33 & 0.04 \\
SN2018dfc & 0.037 & 36.10 & g & 7.5 $\pm$ 0.5 & 9.59 $\pm$ 0.10 & 17.55 & -18.56 & 0.01 \\
 &  &  & r & 10.5 $\pm$ 0.9 & 9.11 $\pm$ 0.20 & 17.60 & -18.50 & 0.02 \\
SN2018dfi & 0.031 & 35.76 & g & 22.7 $\pm$ 0.8 & 3.45 $\pm$ 0.03 & 18.66 & -17.10 & 0.01 \\
 &  &  & r & 25.4 $\pm$ 1.1 & 5.64 $\pm$ 0.20 & 18.12 & -17.64 & 0.04 \\
SN2018egh & 0.038 & 36.17 & g & 8.3 $\pm$ 1.7 & 1.61 $\pm$ 0.04 & 19.48 & -16.69 & 0.03 \\
SN2018fzn & 0.037 & 36.16 & g & 19.7 $\pm$ 1.1 & 2.95 $\pm$ 0.03 & 18.83 & -17.34 & 0.01 \\
 &  &  & r & 23.1 $\pm$ 0.7 & 3.95 $\pm$ 0.06 & 18.51 & -17.65 & 0.02 \\
SN2018fif & 0.017 & 34.43 & g & 12.2 $\pm$ 0.4 & 10.30 $\pm$ 0.01 & 17.47 & -16.97 & $< 10^{-2}$ \\
 &  &  & r & 16.4 $\pm$ 2.7 & 12.80 $\pm$ 0.02 & 17.23 & -17.20 & $< 10^{-2}$ \\
SN2018fsm & 0.040 & 36.30 & g & 6.7 $\pm$ 0.8 & 6.68 $\pm$ 0.06 & 17.94 & -18.37 & 0.01 \\
 &  &  & r & 9.5 $\pm$ 0.6 & 6.08 $\pm$ 0.01 & 18.04 & -18.26 & $< 10^{-2}$ \\
SN2018gts & 0.030 & 35.63 & g & 6.5 $\pm$ 0.7 & 2.74 $\pm$ 0.02 & 18.91 & -16.73 & 0.01 \\
 &  &  & r & 8.6 $\pm$ 0.6 & 4.76 $\pm$ 0.08 & 18.31 & -17.33 & 0.02 \\
SN2018grf & 0.050 & 36.81 & g & 5.0 $\pm$ 0.1 & 4.40 $\pm$ 0.00 & 18.39 & -18.41 & $< 10^{-2}$ \\
 &  &  & r & 7.1 $\pm$ 0.8 & 4.14 $\pm$ 0.03 & 18.46 & -18.35 & 0.01 \\
SN2018gvn & 0.040 & 36.30 & g & 5.0 $\pm$ 0.0 & 4.85 $\pm$ 0.09 & 18.29 & -18.02 & 0.02 \\
SN2018iua & 0.040 & 36.30 & g & 6.4 $\pm$ 1.1 & 2.27 $\pm$ 0.04 & 19.11 & -17.20 & 0.02 \\
 &  &  & r & 15.5 $\pm$ 1.0 & 2.66 $\pm$ 0.00 & 18.94 & -17.37 & $< 10^{-2}$ \\
SN2018leh & 0.024 & 35.17 & g & 13.4 $\pm$ 1.0 & 14.70 $\pm$ 0.00 & 17.08 & -18.08 & $< 10^{-2}$ \\
 &  &  & r & 17.0 $\pm$ 1.0 & 14.80 $\pm$ 0.03 & 17.07 & -18.09 & $< 10^{-2}$ \\
\hline
\hline
\end{tabular}
\label{mag_table}
\end{table*}

\subsubsection{The Flash events}

The identification of flash features in this work is solely based on the study of the spectral range surrounding the strong \ion{He}{2} emission line at $4686$\,\AA. This follows previous work \citep{khazov2016}) and is also supported by large-scale theoretical model grids \citep{grohb2020} that show that this feature is ubiquitous in early spectra ($<2$\,d). We choose not to use hydrogen lines as a marker for flash features since host galaxy lines could contribute to it. 

In previous well-studied cases of events with high-quality early spectra, such as SN 2013fs \citep{yaron2017} and SN 2013cu \citep{galyam2014}, the line \ion{He}{2} $\lambda4686$ is very prominent with a profile that is often well described by a narrow core with broad Lorentzian wings, which could be attributed to electron scattering within the CSM.

\begin{figure*}
\hspace{-1.1cm}
    \includegraphics[trim = 1.5 80 1 100 ,clip,scale = 0.68]{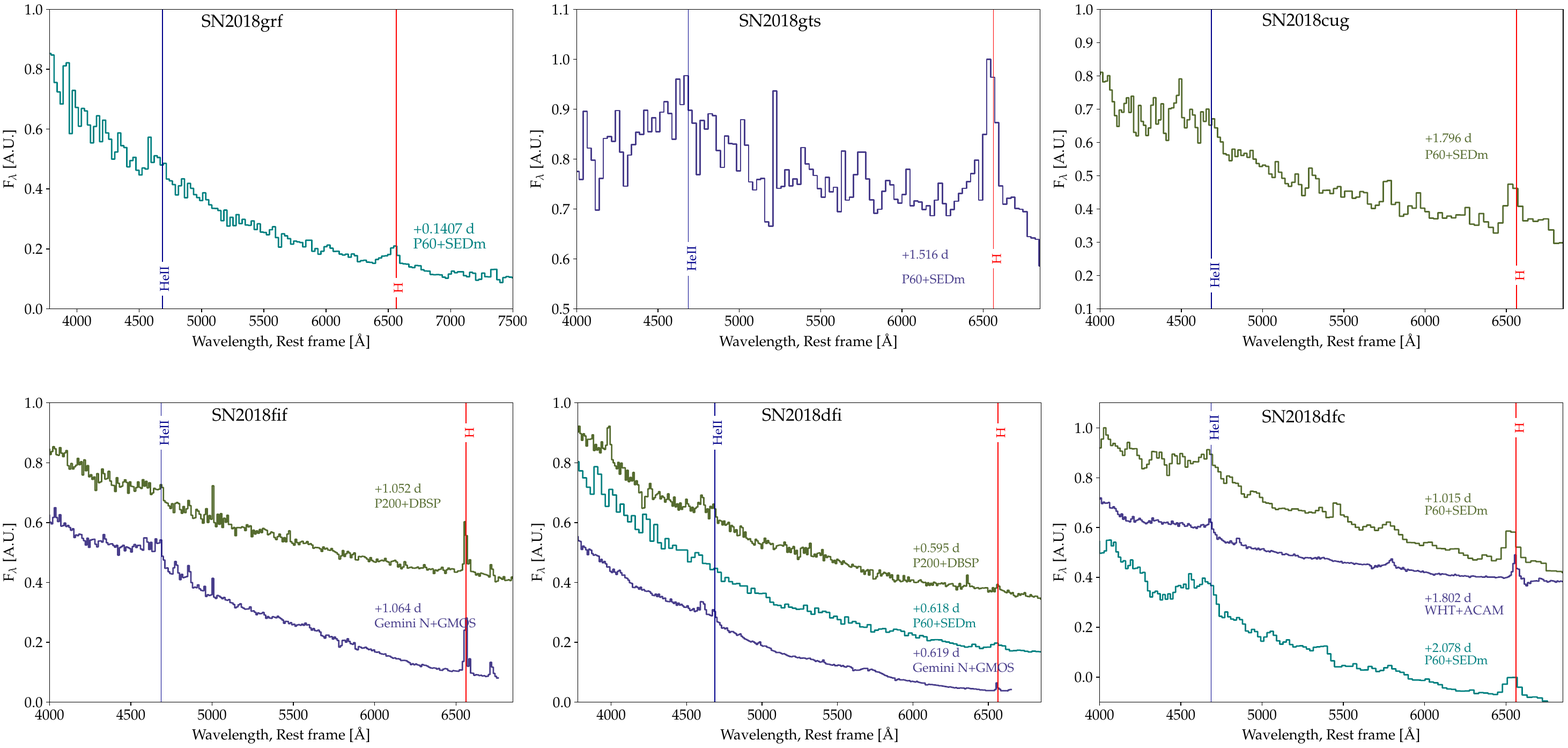}
    \caption{A collection of spectra of six confirmed Flashers. The acquisition time of the spectra are with regard to the estimated explosion date.}
    \label{fig:flashers}
\end{figure*}

As discussed in detail by \cite{soumagnac2019}, as the spectra of such events evolve with time, the strong \ion{He}{2} emission line is replaced by a ledge-shaped feature that is probably composed of blended high-ionization lines of C, N and O. Both the \ion{He}{2} line and the other lines are sometimes detected as a narrow emission line on top of the ledge-shaped feature (see Fig~\ref{fig:flashers} and Fig. 7 of \citealt{soumagnac2019}).

As several of our early spectra were obtained with the low-resolution SEDm instrument (in particular those of SN 2018grf, SN 2018gts and SN 2018cug), we can not easily differentiate between the various manifestations of the excess emission around $4686$\,\AA. We therefore adopt the detection of excess emission around this wavelength as our criterion for defining an object as having flash features. Analysis of the cases where we have both early SEDm spectra as well as high spectral resolution data from larger telescopes (e.g., SN 2018dfc), confirm the nature of the emission we see in the SEDm spectra and support this approach (Fig.~\ref{fig:flashers}).

\begin{figure}
\hspace{-1cm}
    \includegraphics[scale=0.45]{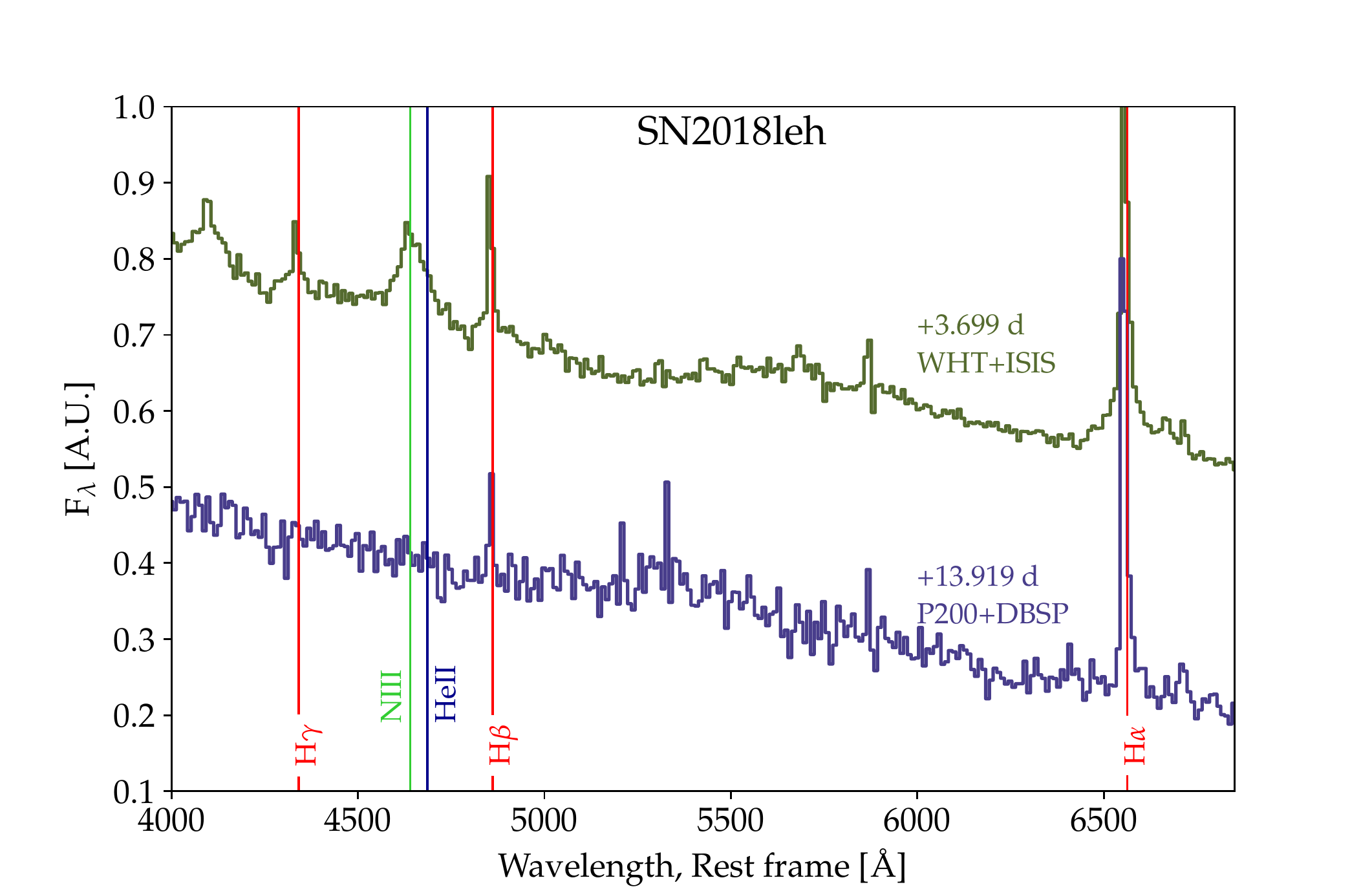}
    \caption{Spectroscopic evolution of SN 2018leh, a Type II SN that shows transient He II emission 4 days after its estimated explosion time.}
    \label{fig:18leh}
\end{figure}

SN 2018leh is the seventh object which displayed flash features. It does not belong to the sub-sample we are considering for this study since its first spectrum was obtained $\approx 3.7$ days after the estimated explosion time. This object shows the Balmer emission lines H$\alpha$, H$\beta$,and H$\gamma$, that persist for an extended period of time, $\approx 10 $ days, which led us to classify this event as a SN IIn. The first spectrum also shows a strong He II line which does not show in the spectrum obtained about 10 days later, see Fig. \ref{fig:18leh}. The transient He II line would technically qualify this event as a member of the flash class. A discussion of the group of objects displaying long-lived flash features and their relation to SNe IIn is outside the scope of this paper.

\subsubsection{The Non-flashers}

\begin{figure*}
    \includegraphics[trim= 5 120 5 150 ,clip,scale=0.6]{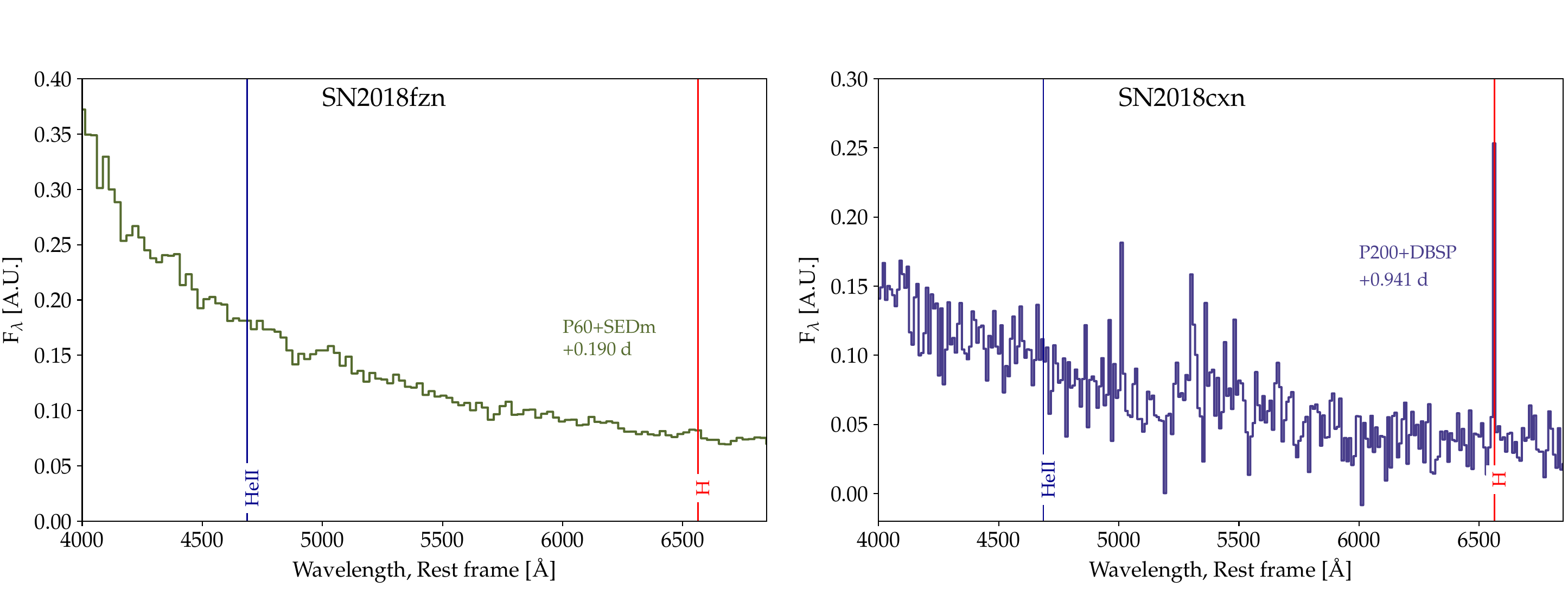}
    \caption{Early spectra of non-flashers SN 2018fzn and SN 2018cxn. These spectra were both obtained within less than a day from the estimated time of explosion. Only a smooth continuum is observed. }
    \label{fig:no_flashers}
\end{figure*}

We consider an event as lacking flash features when we have early, high-quality spectra (i.e. high S/N or higher resolution than SEDm) that do not show any excess emission around \ion{He}{2} $4686$\,\AA. Often, this means that the spectrum is blue and featureless. Among the ten events included in our 2-day sub-sample, SN 2018fzn was observed shortly after explosion (0.19\,d, Table~\ref{Master_table}) with SEDm. While the resolution is low, the signal to noise is sufficient to determine that we cannot find any hint of possible excess emission (Fig.~\ref{fig:no_flashers}). Based on the few previous events with spectra that were obtained so early after explosion (in particular SN 2013fs; \citealt{yaron2017}), we would expect strong emission lines that would be observable with SEDm (see the simulation in Extended Data Figure 2 of \citealt{galyam2014}). The first spectrum of SN 2018cxn was obtained with P200/DBSP less than a day past explosion. The higher resolution and the complete absence of He II emission (Fig.~\ref{fig:no_flashers}) suggest that there were no flash features. In both cases, we conclude that there are no indications for a circumstellar shell. 

\subsubsection{The dubious flashers}

\begin{figure}
    \centering
    \includegraphics[trim = 15 40 30 90, clip, scale = 0.75]{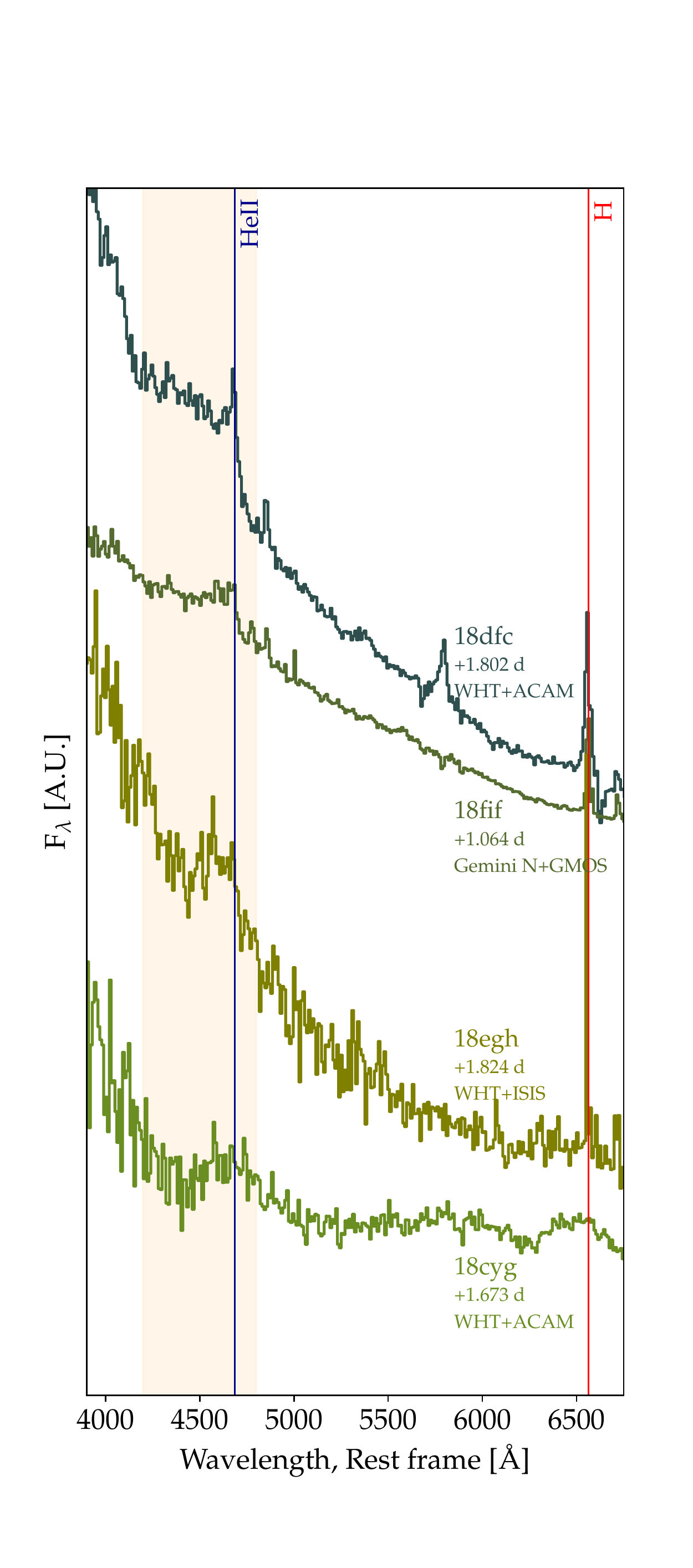}
    \caption{Candidates showing a wide bump-like structure close to the He II emission line. We highlight in orange the region we searched for excess emission.}
    \label{fig:bumpers}
\end{figure}

SN 2018cyg and SN 2018egh both show excess flux around $4686$\,\AA\, (Fig.~\ref{fig:bumpers}). However, this excess does not resemble the ledge-shaped feature seen for example in the spectra of SN 2018fif \citep{soumagnac2019}, and discussed above. An additional complication is that the spectra of SNe II at early phase (prior to the appearance of strong and broad hydrogen Balmer lines) sometime show an absorption complex extending between $\approx4000-4500$\,\AA~  (E. Zimmerman et al., in preparation). Such a complex appears in the spectra of both SN 2018cyg and SN 2018egh. It is difficult to determine whether the apparent bump around $4600$\,\AA\, represents an actual excess, or if it rather is the continuum edge redward of an absorption feature. In addition, even though we have secured early, high resolution spectra for these objects (Table~\ref{Master_table}) they both lack a narrow emission component from He~II. These broad features are however transient and do not appear at later times. These issues makes it difficult to determine whether these events show flash features or not.

As an additional test of whether these two objects show a flux excess around $4600$\,\AA, we conducted the following test: we constructed model spectra composed of black body continua, over which we superpose model Gaussian emission lines whose width is a free parameter (with typical best fits of $\approx 100$\,km\,s$^{-1}$), in those cases (in particular, SN 2018dfc) where such lines are apparent. In addition, we add 
 a broad feature extending between $4200-4750$\,\AA, which we defined by fitting a third order polynomial to the ledge-shaped feature appearing in the SN 2018fif WHT spectrum (Fig.~\ref{fig:bumpers}). The data was fitted using the python package \textit{iminuit} \citep{iminuit}. We then performed a $\chi^2$ test to determine whether the bump feature is significantly detected (in the sense that $\Delta\chi^2>1$ between models) when comparing the goodness of fit over the intervals given in Table~\ref{tab:bump_models}. 
 
\begin{figure*}
    \hspace{-1.5cm}
    \includegraphics[trim = 1 120 2 120, clip,scale= 0.7]{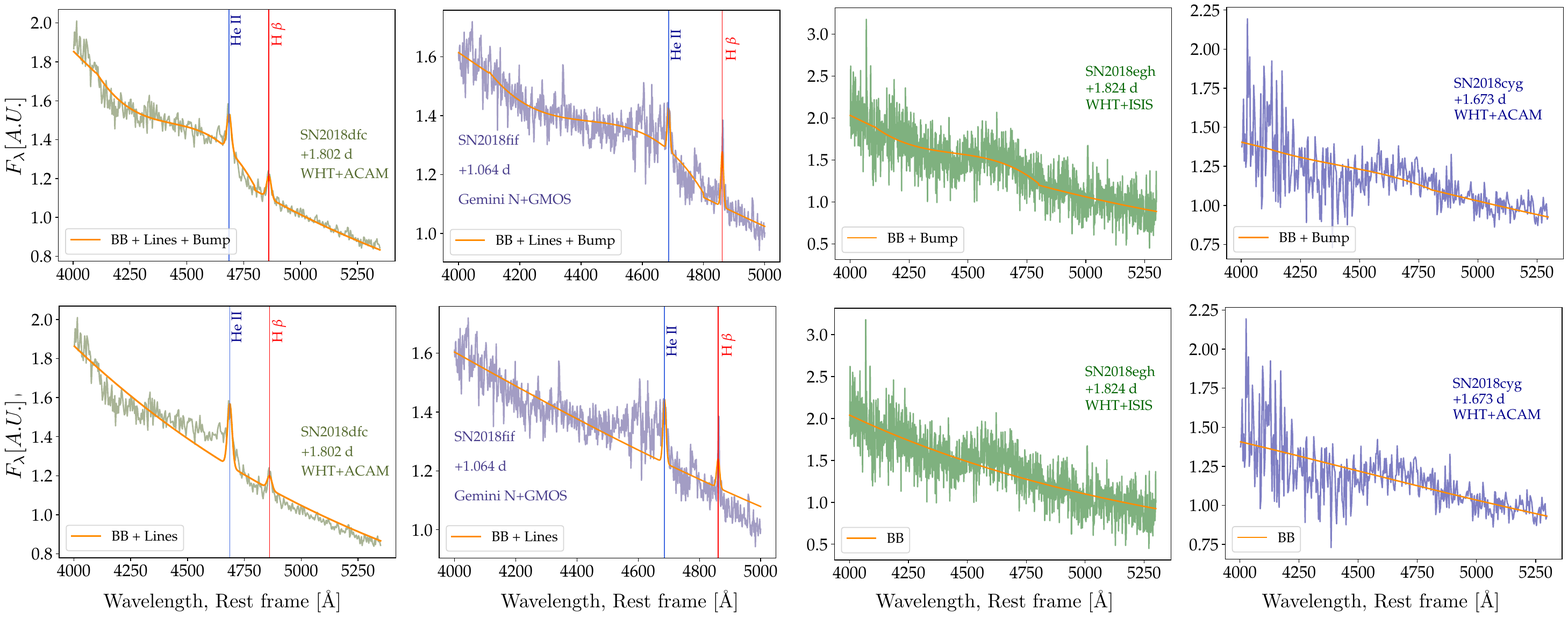}
    \caption{Fit results with (top panels) and without (bottom panels) the broad feature component for SNe 2018dfc, 2018fif, 2018egh and 2018cyg (from left to right). No narrow emission lines are seen in the spectra of 2018egh and 2018cyg, and neither provides a significant detection of a bump component.}
    \label{fig:model_nobump}
\end{figure*}

The results of these model comparisons are reported in Table~\ref{tab:bump_models} and Figure~\ref{fig:model_nobump}. As can be seen, the bump is strongly detected in the spectra of SN 2018dfc (and is also recovered for SN 2018fif), but neither for SN 2018cyg nor SN 2018egh. The results do not change if we try to also fit narrow lines even to spectra where no obvious lines are seen, or if we try to fit additional weaker line features such as H$\gamma$. For SN 2018dfc, the bump feature is detected both in the earlier low-resolution SEDm spectrum (at low significance) and clearly in the later high-resolution WHT spectrum. We therefore conclude that we can not ascertain that SN 2018cyg and SN 2018egh show flash features. We conduct our analysis below and report our results for all possible options (i.e., that both, one, or neither of these show evidence for CSM). 

\begin{table*}
    \centering
    \caption{Results of test fits for models with and without the broad bump feature.}
    \begin{tabular}{lllllll}
    \hline
        Name & Spectrum & Lines fit & $\chi_{2}$/dof  & $\chi_{2}$/dof  & Fit Interval \\
        &&&with bump&without bump&[Å]\\
        \hline
        \hline
        SN 2018dfc & P60+SEDm +1.015\,d & [HeII, H$\beta$] & 0.76 & 1.43 & 4000-5300 \\
        SN 2018dfc & WHT+ACAM +1.082\,d & [HeII, H$\beta$] & 1.66 & 4.09 & 4000-5300 \\
        SN 2018fif & Gemini+GMOS +1.064\,d & [HeII, H$\beta$] & 2.12 & 3.34 & 4000-5000 \\
        SN 2018egh & WHT+ISIS +1.824\,d & [HeII, H$\beta$] & 0.87 & 0.91 & 4000-5300 \\
        SN 2018egh & WHT+ISIS +1.824\,d & No Lines & 0.87 & 0.93 & 4000-5300 \\
        SN 2018cyg & WHT+ACAM +1.673\,d & No Lines & 0.90 & 0.90 & 4000-5300 \\
        
        \hline
 
    \end{tabular}
    \vspace{1cm}
    \label{tab:bump_models}
\end{table*}


\vspace{2cm}
\section{Discussion and conclusions}

\subsection{How common are flash features}

\begin{figure}
    \centering
    \includegraphics[trim = 1 2 2 2, clip,scale= 0.4]{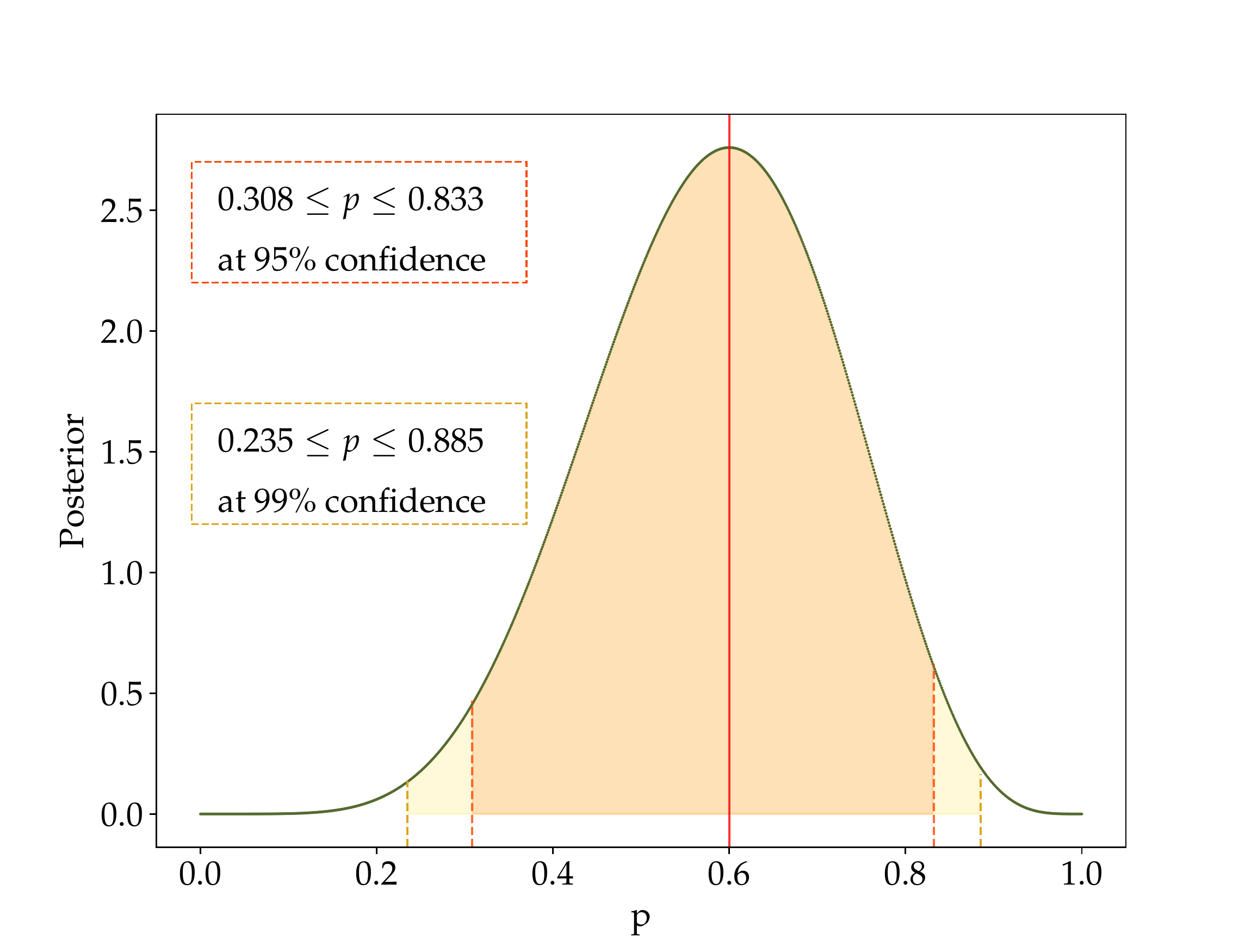}
    \caption{Posterior probability distribution vs. the probability to observe a flash ionised event. This analysis is based on the subsample of infant candidates which had a first spectrum within $< 2$ days from the estimated explosion date. The lower limit is 30.8\% (23.5\%) at 95\%(99\%) confidence interval.}
    \label{fig:flash_stat}
\end{figure}

Based on our systematic survey of infant SNe II with spectra obtained within two days of discovery, we have found that at least $60\%$, and perhaps as many as $80\%$ of the sample of ten events show evidence for flash-ionized emission. Taking into account our limited sample size and assuming binomial statistics ($\mathcal{B}(k,n,p)$), we infer the true fraction of SNe with CSM that manifests as flash features from the true probability p to observe a flash event given we the observed fraction D using a Bayesian model: 

\begin{equation}
    P(p|D) = \frac{P(D|p)\times \pi(p)}{P(D)}
\end{equation}
Where $p$ is the probability of observing a flash ionised event (here $p \in [0;1]\,$), D is the observation presented in this paper (i.e.: 6 out of 10 candidates are showing flash features). The probability of our observation, $P(D)$ can be calculated with the formula of total probability, i.e. $P(D) = \int_0^1 \mathcal{B}(6,10,p)\times\pi(p)\, \mathrm{d}p $. We assume a uniform distribution for the prior $\pi(p)$ which allows us to write the posterior function as: 
\begin{equation}
    P(p|D) = \frac{ {{10}\choose{6}} p^6 (1-p)^4 }{\int_{0}^1 {{10}\choose{6}} p^6 (1-p)^4 \mathrm{d}p}
\end{equation}
This results in a Beta distribution (see Figure \ref{fig:flash_stat}). We can put a strict lower limit on the fraction of infant SNe II showing flash features of $>30.8\%$ ($>23.5\%$) at the $95\%$ ($99\%$) confidence level. This fraction rapidly drops when events with spectra obtained within $7$\,d from explosion are considered; presumably the fraction could be even higher for events with even earlier spectra.  

These results are broadly consistent with previous work by \cite{khazov2016}, which estimate that $7-36\%$ show flash features in spectra obtained within $<2$\,d from explosion ($68\%$ confidence level). It is also consistent with the low observed frequency of flash features among the general population of Type II SNe reported in the literature, as these events very rarely have a spectrum obtained $<2$\,d after explosion, and Table~\ref{Master_table} shows that the fraction of flash events falls rapidly at ages $>2$\,d. The unique nightly cadence of the ZTF partnership survey enabled us to routinely discover infant SNe and rapidly obtain spectra, while the systematic design of our survey allowed for a robust measurement of the frequency of this phenomenon. 

\subsection{Possible biases}

\begin{figure*}
    \centering
    \includegraphics[trim = 90 60 110 60 , clip,scale = 0.8]{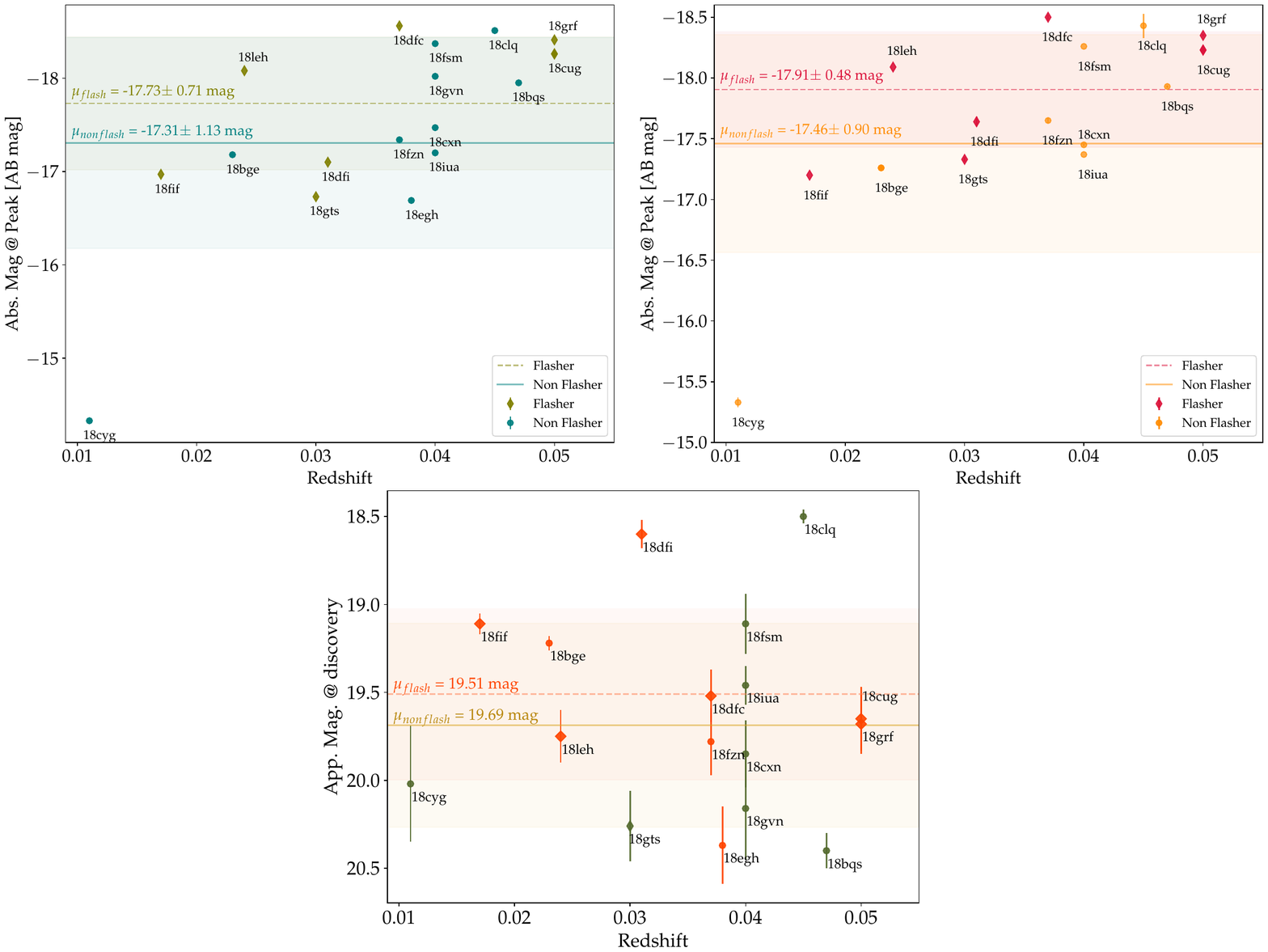}
    \caption{Top: absolute magnitude in r band (left) and g band (right) vs. redshift.  Bottom: apparent magnitude at discovery vs. redshift. Color bands represent the error on the mean peak magnitude for both flash and non flash groups. SN 18cyg is host reddened and hence appears very faint, see text.}
    \label{fig:mag_disco_peak}
\end{figure*}

\cite{khazov2016} (see their Fig. 8) show that Type II SNe showing flash-ionized features tend to be brighter at peak than other events. We cannot confirm this is also true for our sample. We consider here the subsample of Infant Supernovae whose first spectrum was obtained within less than 7 days from the estimated explosion time, the peak magnitudes were obtained following the method described in § 3.3.2. Figure ~\ref{fig:mag_disco_peak}, top panel, shows the peak magnitudes in both g and r bands for flashers and non flashers. Flashers appear to be brighter in both bands. However, when one considers SN 2018cyg as a flasher, the average peak magnitude of both groups is inverted and non-flashers appear brighter than flashers (see Table \ref{tab:peakmags}, top section). Since SN 2018cyg is strongly reddened, we repeated this same analysis but with SN 2018cyg being host extinction corrected. To apply the extinction correction, we consider the spectrum from 2018 August, 4 \footnote{see on WISeREP : https://wiserep.weizmann.ac.il/object/698 } and apply the method described in \citealt{Poznanski_2012}, using the line doublet of sodium. We consider the doublet not to be resolved and apply the following formula: 
\begin{equation}
    \log_{10}(E_{B-V}) = 1.17 \, \times EW(D_1+D_2)-1.85 \pm 0.08
\end{equation}
We estimate the EW of D$_1$+D$_2$ using the built-in tool from WISeREP by measuring it several times. The mean EW is 1.64 Å with an error of 0.17 Å. Following Eq. (5), the final peak magnitudes for SN 2018cyg are : $M_{peak,r} = -18.45 \pm 0.50$ and $M_{peak,g} = -18.77 \pm 0.80$. Table \ref{tab:peakmags} summarises the different cases : whether SN 2018cyg is a flasher and whether SN 2018cyg was corrected for estimated host extinction. We find that flash events are not inherently brighter than non-flash events. 

\begin{table*}
 \caption{Peak magnitude comparison between the flash events and the non flash events. }
\textbf{r band}
\centering
    \begin{tabular}{ccc}
    \toprule
        & M$_{peak,\,flasher}$  & M$_{peak,\,non\,flasher}$  \\
        \hline
        \textit{18cyg not corrected for extinction}\\
        \hline
        18cyg $\subset$ flasher & $-17.58 \pm 0.96 $  & $-17.76 \pm 0.42$ \\
        18cyg $\not\subset$ flasher & $-17.91 \pm 0.48$ & $-17.46 \pm 0.90$ \\
        \hline
        \textit{18cyg corrected for extinction}\\
        \hline
        18cyg $\subset$ flasher & $-17.97 \pm 0.48 $  & $-17.76 \pm 0.42$ \\
        18cyg $\not\subset$ flasher & $-17.91\pm 0.48$ & $-17.85 \pm 0.46$ \\
        \hline
        \hline
    \end{tabular}  \\
\vspace{5pt}
\textbf{g band}
    \begin{tabular}{ccc}
        \toprule
         & M$_{peak,\,flasher}$ & M$_{peak,\,non\,flasher}$ \\
        \hline
        \textit{18cyg not corrected for extinction}\\
        \hline
        18cyg $\subset$ flasher & $-17.30 \pm 1.31 $ & $-17.64 \pm 0.57$\\
        18cyg $\not\subset$ flasher &$-17.73 \pm 0.71$ & $-17.31 \pm 1.13$  \\
        \hline
        \textit{18cyg corrected for extinction}\\
        \hline
        18cyg $\subset$ flasher &$-17.86 \pm 0.75 $ & $-17.64 \pm 0.57$\\
        18cyg $\not\subset$ flasher &$-17.76 \pm 0.75$ & $-17.75 \pm 0.64$  \\
        \hline
        \hline

    \end{tabular}
    \tablecomments{ This analysis is performed with the subsample which has a first spectrum within less than seven days from the estimated explosion time.}
    \label{tab:peakmags}
\end{table*}


We also inspect in Fig.~\ref{fig:mag_disco_peak} (lower panel) the distribution of apparent magnitudes at discovery for our $<7$\,d sample. As can be seen there, we find that the flash events were not significantly brighter at discovery than other events, and thus neither more likely to be discovered, nor to be followed, as both of these depend on the apparent magnitude of the object at discovery. 

\subsection{Implications}

We have shown here that a significant fraction, and possibly most, Type II SN progenitors, show transient emission lines in their early spectra, that provide evidence that these stars are embedded in a compact distribution of CSM \citep{yaron2017}. The narrow width of these emission lines indicates a slow expansion speed for the CSM ($100-800$\,km\,s$^{-1}$, \citealt{grohb2020} ), and combined with its compact radial dimension ($<10^{15}$\,cm) we have evidence that the CSM was deposited by the stars within months to a few years prior to its terminal explosion.
Assuming these progenitors are mostly red supergiants (RSGs; \citealt{smartt2015}), this would suggest that most exploding RSGs experience an enhanced mass loss shortly prior to explosion.

While RSGs certainly lose mass during their final stages of evolution \citep{smith2014}, such a period of enhanced mass loss shortly (months to a year) prior to explosion is not explained by standard stellar evolution models. Our work thus may indicate that additional physical processes leading to such pre-explosion instabilities (e.g., \citealt{Arnett2011}, \citealt{Shiode2014}) not only exist, but are ubiquitous among massive stars.

As we have shown that most SN II progenitors likely undergo a remarkable evolution shortly prior to explosion, it may be needed to re-examine the stellar models used as initial conditions to explosion simulations. In particular, at least some of the effects proposed to explain such pre-explosion mass loss, may render the spherical pre-explosion stellar models used in explosion simulations less realistic \citep{Arnett2016}. Perhaps our work thus provides a clue how to tackle some of the problems encountered in trying to reproduce the observed distribution of SN explosions using numerical explosion models. 

\section{Conclusions}
We report the results from the first year (2018) of our systematic survey for infant Type II SNe in the ZTF partnership survey. We collected 28 such objects (at a rate of about one per week) and obtained rapid follow-up spectroscopy within $2$\,d from explosion for 10 events.
Between $6-8$ of these show evidence for transient emission from a surrounding distribution of CSM, and we can thus place a strict lower limit of $>30\%$ (at $95\%$ C.L.) on the fraction of SN II progenitors that explode within compact CSM distributions. This finding is inconsistent with predictions from standard stellar evolution models, and suggests that additional physics is required to explain the final stages ($\sim1$\,year prior to explosion) of massive star evolution. The structural changes that may accompany such final episodes of intense mass loss can modify the stellar structure prior to explosion and may require adjusting the initial conditions assumed for core-collapse SN explosion simulations, and may thus shed light on the yet unsolved question of how massive stars end their life in supernova explosions.

\section{Acknowledgements}

AGY's research is supported by the EU via ERC grant No. 725161, the ISF GW excellence center, an IMOS space infrastructure grant and BSF/Transformative and GIF grants, as well as The Benoziyo Endowment Fund for the Advancement of Science, the Deloro Institute for Advanced Research in Space and Optics, The Veronika A. Rabl Physics Discretionary Fund, Paul and Tina Gardner, Yeda-Sela and the WIS-CIT joint research grant;  AGY is the recipient of the Helen and Martin Kimmel Award for Innovative Investigation.
The ztfquery code was funded by the European Research Council (ERC) under the European Union's Horizon 2020 research and innovation programme (grant agreement No. 759194 - USNAC, PI: Rigault).
The ZTF forced-photometry service was funded under the Heising-Simons Foundation grant \#12540303 (PI: Graham). Based on observations obtained with the Samuel Oschin 48-inch Telescope at the Palomar
Observatory as part of the Zwicky Transient Facility project. ZTF is supported by the National
Science Foundation under Grant No. AST-1440341 and a collaboration including Caltech, IPAC,
the Weizmann Institute for Science, the Oskar Klein Center at Stockholm University, the University
of Maryland, the University of Washington, Deutsches Elektronen-Synchrotron and Humboldt
University, Los Alamos National Laboratories, the TANGO Consortium of Taiwan, the University
of Wisconsin at Milwaukee, and Lawrence Berkeley National Laboratories. Operations are
conducted by COO, IPAC, and UW. The data presented here were obtained [in part] with ALFOSC, which is provided by the Instituto de Astrofisica de Andalucia (IAA) under a joint agreement with the University of Copenhagen and NOTSA.  A.A.M.~is funded by the LSST Corporation, the
Brinson Foundation, and the Moore Foundation in support of the LSSTC Data
Science Fellowship Program; he also receives support as a CIERA Fellow by the
CIERA Postdoctoral Fellowship Program (Center for Interdisciplinary
Exploration and Research in Astrophysics, Northwestern University).

\section{Appendix}

The full list of candidate infant SNe II returned by {\tt ztfquery} (see $\S~$\ref{sec:sample}) is given in Table~\ref{tab:infants}. 
Of the 43 candidates, inspection shows that 15 are spurious, and these have been removed from out sample. We provide some comments on removed objects.

\paragraph{Early false positives} A group of objects detected right at the start of the survey (during March 2018 till early April) suffered from unreliable photometry, manifest as a mix of detections and non-detections during the same period, and often during the same night. This is likely due to problematic early references. The mix of detections and non-detections created artificial triggers due to a spurious non-detection just prior to the first detection. This group includes ZTF18aaayemw, ZTF18aaccmnh, ZTF18aagrded (which was also detected by ATLAS 3 days prior to the ZTF false non-detection, and reported to the TNS as AT2018ahi), ZTF18aahrzrb, ZTF18aainvic, and ZTF18aaogibq.

\paragraph{ZTF18aaqkdwu} This trigger resulted from a spurious photometry point generated by the pipeline at the location of SN 2019eoe a year prior to the explosion of the actual SN. 
\paragraph{ZTF18aasxvsg} Additional analysis recovered several clear detections prior to the spurious non-detection that triggered this event.

\paragraph{ZTF18abcqhgr} This event is likely a real infant SN II, but we could not recover it using the forced photometry pipeline and it was therefore removed from the sample. This object does not have an early spectrum. 

\paragraph{ZTF18acbwvsp} This event was detected by SNHunt and reported to the TNS as AT 2018hqm a few days prior to the only ZTF non-detection, indicating it is likely not a RI SN. 

\paragraph{ZTF18acecuxq} The early photometry of this event shows a mix of detections and non-detections during the same nights, and was deemed unreliable. A spectrum obtained within a day of the false non-detection (A. Tzanidakis, in preparation) is that of an old SN II, supporting this conclusion. 

\paragraph{ZTF18acgvgiq} This event was detected by ATLAS and reported to the TNS as SN 2018fru more than 2 months prior to the ZTF non-detection, indicating our non-detections preceding the ZTF first detection were spurious.

\paragraph{ZTF18acefuhk} Updated photometry does not recover a non-detection prior to first detection that satisfies our criteria. This object does not have early spectra.

\paragraph{ZTF18acqxyiq} The forced photometry pipeline did not recover the non-detection by the real-time pipeline, leaving the explosion time poorly constrained.

\paragraph{ZTF18adbikdz} This object was detected by Gaia and reported to the TNS as AT2017isr over a month prior to the first detection by ZTF (when it was already declining). Our single non-detection is spurious.

\bibliography{infant_sne_bibliography}

\begin{table*}[h]
\centering
\caption{Results of the search for infant SN II using {\tt ZTFquery} }
\begin{tabular}{lllllll}
Name & RA & Dec & Redshift & First Detection & First spectrum & Real?\\
 & [deg] & [deg] &  & [days] & [days] & \\
\hline
\hline
ZTF18aaayemw & 134.8982936 & 45.6116267 & 0.052 & 2458156.7621 & 0.024 & \xmark\\
ZTF18aaccmnh & 194.9769678 & 37.8589965 & 0.0356 & 2458184.8604 & 0.018 & \xmark\\
ZTF18aagrded & 209.8414748 & 46.0317554 & 0.047 & 2458198.8809 & 0.011 & \xmark\\
ZTF18aahrzrb & 181.397224 & 34.3888035 & 0.04 & 2458217.7371 & 1.001 & \xmark\\
ZTF18aainvic & 256.5204624 & 29.6683607 & 0.03175 & 2458218.9088 & 0.019 & \xmark\\
ZTF18aaogibq & 253.5409858 & 24.721127 & 0.037 & 2458231.8783 & 0.020 & \xmark\\
ZTF18aaqkdwu & 199.7588529 & 45.0263019 & 0.06037 & 2458243.677 & 0.001 & \xmark\\
ZTF18aaqkoyr & 166.0666639 & 50.0306275 & 0.023 & 2458243.6854 & 1.036 & \cmark\\
ZTF18aarpttw & 247.2599041 & 43.6268239 & 0.047 & 2458246.822 & 1.001 & \cmark\\
ZTF18aarqxbw & 276.4265403 & 34.6584885 & 0.048 & 2458246.8404 & 1.878 & \cmark\\
ZTF18aasxvsg & 217.1290246 & 37.0678367 & 0.0248 & 2458244.8361 & 0.018 & \xmark\\
ZTF18aatlfus & 257.1764284 & 28.5206128 & 0.0451 & 2458249.8534 & 1.913 & \cmark\\
ZTF18aavpady & 273.0031098 & 44.3602114 & 0.047 & 2458257.8452 & 0.870 & \cmark\\
ZTF18aawyjjq & 263.0587448 & 36.0740074 & 0.04 & 2458263.796 & 0.011 & \cmark\\
ZTF18aayxxew & 197.1395703 & 45.9861525 & 0.061 & 2458278.7043 & 1.961 & \cmark\\
ZTF18abcezmh & 269.4519011 & 40.0764001 & 0.057 & 2458288.7881 & 0.874 & \cmark\\
ZTF18abckutn & 237.0269066 & 55.7148077 & 0.0401 & 2458290.6992 & 0.834 & \cmark\\
ZTF18abcptmt & 267.3298968 & 49.4124315 & 0.05 & 2458291.7869 & 0.878 & \cmark\\
ZTF18abcqhgr & 254.818188 & 60.4317998 & 0.070396 & 2458291.8048 & 0.021 & \xmark\\
ZTF18abdbysy & 233.5352962 & 56.6968517 & 0.01127 & 2458295.7208 & 0.016 & \cmark\\
ZTF18abddjpt & 278.7048393 & 38.2987246 & 0.07 & 2458295.7913 & 0.021 & \cmark\\
ZTF18abeajml & 252.0323502 & 24.3041089 & 0.03651 & 2458303.7989 & 1.002 & \cmark\\
ZTF18abffyqp & 252.7086818 & 45.397907 & 0.031302 & 2458307.6862 & 0.864 & \cmark\\
ZTF18abgqvwv & 254.3164613 & 31.9632993 & 0.0377 & 2458313.7295 & 0.891 & \cmark\\
ZTF18abgrbjb & 274.9986631 & 51.7965471 & 0.03 & 2458313.7492 & 0.032 & \cmark\\
ZTF18abimhfu & 240.1422651 & 31.6429838 & 0.05 & 2458320.6667 & 0.912 & \cmark\\
ZTF18abojpnr & 297.4871203 & 59.5928266 & 0.0375 & 2458351.7166 & 0.021 & \cmark\\
ZTF18abokyfk & 2.3606444 & 47.3540929 & 0.017189 & 2458351.8659 & 0.887 & \cmark\\
ZTF18abrlljc & 253.1840255 & 70.0882366 & 0.05 & 2458359.7 & 0.054 & \cmark\\
ZTF18absldfl & 33.5997507 & 30.811929 & 0.04 & 2458363.8793 & 0.913 & \cmark\\
ZTF18abufaej & 4.4825733 & 12.0916007 & 0.0625 & 2458368.8738 & 0.036 & \cmark\\
ZTF18abvvmdf & 249.1975409 & 55.7358424 & 0.029597 & 2458375.7154 & 0.016 & \cmark\\
ZTF18abwlsoi & 261.8976711 & 71.5302584 & 0.05 & 2458377.6334 & 0.895 & \cmark\\
ZTF18abyvenk & 273.9764532 & 44.6964862 & 0.04 & 2458385.6212 & 0.858 & \cmark\\
ZTF18acbwvsp & 341.9067649 & 39.8806077 & 0.017062 & 2458423.6368 & 0.907 & \xmark\\
ZTF18acecuxq & 68.8323442 & 17.1948085 & 0.02572 & 2458431.8168 & 1.011 & \xmark\\
ZTF18acefuhk & 136.7936282 & 43.9207446 & 0.057 & 2458426.9469 & 0.951 & \xmark\\
ZTF18acgvgiq & 204.0157722 & 66.3012068 & 0.01055 & 2458432.0181 & 1.966 & \xmark\\
ZTF18achtnvk & 96.1687142 & 46.5039037 & 0.04 & 2458434.9036 & 0.043 & \cmark\\
ZTF18acploez & 130.03737 & 68.9031912 & 0.04 & 2458440.9658 & 1.957 & \cmark\\
ZTF18acqxyiq & 149.8258285 & 34.895493 & 0.03849 & 2458443.9437 & 0.001 & \xmark\\
ZTF18adbikdz & 252.014493 & 26.2118328 & 0.03432 & 2458482.0504 & 0.004 & \xmark\\
ZTF18adbmrug & 61.2637352 & 25.2619198 & 0.02396 & 2458482.6991 & 1.897 & \cmark\\
\hline
\end{tabular}
\tablecomments{43 candidates were found, of which 15 ($\sim35\%$) were spurious, leaving 28 infant SNe II in our sample}
\label{tab:infants}
\end{table*}

\end{document}